\documentclass[aps,twocolumn,prc,superscriptaddress,showpacs,nofootinbib,floatfix,amssymb,amsfonts,amsmath]{revtex4-2}
\usepackage{graphicx}% Include figure files
\usepackage{dcolumn}% Align table columns on decimal point
\usepackage{bm}% bold math
\usepackage{xcolor}
\usepackage{amsmath}    % need for subequations
\usepackage{amsfonts}   %note how statements can be commented out
\usepackage{amssymb}
\usepackage{graphicx}   % for figures
\usepackage{multirow} 

\begin{document}

\title{Basis truncation, statistical errors, and systematic uncertainties in relativistic approaches to nuclear response}

\author{A.\ V.\ Afanasjev}
\affiliation{Department of Physics and Astronomy, Mississippi
State University, MS 39762}

\author{E.\ Litvinova}
\affiliation{Department of Physics, Western Michigan
University, MI 49009}

\author{B.\ Osei}
\affiliation{Department of Physics and Astronomy, Mississippi
State University, MS 39762}

\date{\today}

\begin{abstract}
Although there exists a clear and, in principle, exact theoretical formulation for the equation of motion for the response of a correlated fermionic system, its numerical implementations for atomic nuclei require feasible approximations. One of the widely accepted approximations is a truncated harmonic oscillator (HO) basis, whose wave functions are used to expand the solutions obtained with realistic interactions. In this work, we extend previously employed HO basis truncated at $N_F$ = 20 fermionic shells to $N_F$ = 50 and perform a systematic study of the effects of such basis increase on nuclear resonances.
The relativistic random phase approximation (RRPA) and its extension by the particle-vibration coupling dubbed as relativistic time-blocking approximation (RTBA) are applied to the description of
the monopole, dipole, quadrupole, and octupole resonances in $^{48}$Ca, $^{78}$Ni, and $^{132}$Sn, and the RRPA studies are extended to $^{70}$Ca and $^{208}$Pb. A considerable sensitivity of the strength distributions to the HO basis size is found, especially for low-spin
resonances in the light neutron-rich nuclei.  The effects of the HO basis extension to $N_F$ = 50 are analyzed and linked to the involvement
of proton and neutron continuum states and proton quasi-bound states in the strength formation. The obtained results point to the importance of the HO basis completeness and continuum effects 
in the nuclear response calculations and evaluation of the associated parameters of the nuclear equation of state.
Statistical errors and systematic uncertainties in the RRPA strength functions are analyzed.
They are found to be substantial for the monopole response, but significantly smaller for the dipole, 
quadrupole, and octupole ones.  Neither of them shows a pronounced mass dependence, and 
statistical errors are generally smaller than systematic uncertainties. 

\end{abstract}

\maketitle

%%%%%%%%%%%%%%%%%%%%%%%%%%%%
\section{INTRODUCTION}
%%%%%%%%%%%%%%%%%%%%%%%%%%%%%

Knowledge about the nuclear response is critical for understanding a variety of nuclear structure 
phenomena from strong interactions in correlated media to electromagnetic and weak decays in stars.
The response functions and associated strength distributions in various spin-parity ($J^{\pi}$), and 
isospin ($T$) channels probe different facets of nuclear forces and quantify the spectral properties, 
namely, the energies of the nuclear excited states characterized by the $\{J^{\pi},T\}$ quantum 
numbers and the probabilities of their excitations. Most commonly, excitations between the ground 
and excited states are treated by the response formalism, although generalizations to transitions between 
excited states are also possible.  Nuclear spectra constitute inputs to many applications, from nuclear 
technologies to fundamental physics frontiers, and only a part of the required spectral strength distributions 
in limited areas of the nuclear landscape is accessible experimentally. Theory serves to provide the 
remaining input, which is often required to be compatible with high accuracy standards and quantified 
uncertainties.  

There are several sources of theoretical uncertainty in the description of atomic nuclei, which accumulate due to the fact that nuclear structure phenomenology is sensitive to several energy scales that can not be completely decoupled within the effective theory frameworks. For instance, quantum hadrodynamics (QHD) is not continuously calculable from quantum chromodynamics, and the strongly-correlated nuclear medium is not satisfactorily calculable from the QHD without introducing new low-energy constants at a particular scale. Furthermore, even effective theories at the given low-energy scale of nuclear structure admit various parametrizations. A famous example is the nuclear density functional theory (DFT) based on the approximate reduction of the many-body problem to the one-body problem. Various forms of the energy density functionals (EDFs) exist, and even different parametrizations of the same ans\"atze 
are in active use 
by practitioners \cite{VALR.05,Colo2020}.  

  The approximations to the many-body problem beyond the mean field set the next class of uncertainties, 
which, for the case of nuclear response, can be defined and organized by configuration complexity \cite{LitvinovaSchuck2019,Novak2024}.  While the completeness of the response theory in terms of 
complex configurations is still in progress, it becomes evident that on each complexity level,  the 
uncertainties associated with numerical implementations are inevitable.

In view of the universality and broad applications of the nuclear response theory, it is important to 
quantify theoretical uncertainties  of its practical implementations for various multipolarities across 
the nuclear chart. The present paper aims to define such uncertainties  related to  (i) the truncation 
of the harmonic oscillator (HO) basis used for expanding the nucleonic wave functions of the 
self-consistent mean field (utilized, in turn, as a complete basis for the response calculations),  
(ii) statistical errors emerging from the details of the fitting protocol of the employed functional, and  
(iii) systematic  uncertainties emerging from the definition of the functionals. 
 
   The first goal of this study is motivated by the fact that the majority of nuclear response calculations are carried out in frameworks that employ the HO basis set expansion for the 
nucleonic wave functions.  Because of numerical reasons, this basis is truncated.  For example, in the DFTs, a finite fermionic basis 
characterized by the number $N_F$ of full fermionic  shells is used in the calculations (see, for 
example, Refs.\ \cite{DIRHB-code.14,TOAPT.24}). In static calculations, the truncation of the basis 
is dictated by the convergence of the total binding energies of the ground states of the nuclei under 
study, and recent studies allow the assessment of the accuracy of such truncation with respect to 
numerically exact results obtained in an infinite fermionic basis \cite{NL5Z-DDMEZ-PCZ,OAD.25}.

The same convergence criteria are used in the truncation of the HO basis for the nuclear response 
calculations:  for example, $N_F=18$ and $N_F=20$ bases have been used in relativistic and non-relativistic 
calculations of the nuclear response in recent years (see, for example, Refs.\ \cite{VTCDMP.12,PhysRevC.111.054314,LitvinovaRingTselyaev2007,LitvinovaSchuck2019}).
However, the detailed analysis presented in Sec.\ \ref{trunc-basis-sp} of the present
paper reveals a substantial difference in the convergence of single-particle states with
negative and positive energies. The convergence of the former is defined by the convergence 
of total binding energies, and it is reached at $N_F\approx 20$ for most of the cases.
In contrast, this is not a case for positive energy quasi-bound and continuum states, 
the convergence of which is defined by hard-wall boundary conditions in coordinate
space effectively imposed by the use of a finite HO basis, similarly to nuclear 
many-body calculations of Refs. \cite{FHP.12,BEHPW.16}.  It turns out that the increase of
$N_F$ leads to a lowering of the energies of such states and densification of the quasi-continuum. 
This immediately raises the question about the sensitivity of the properties of excited states 
to the size of the fermionic basis dependent on $N_F$.

The energies and wave functions of the single-particle states are the major building blocks for the description of nuclear excitations within any framework, including the response theory. This sets the first subject for the present investigation, which links the changes in the single-particle spectrum due to the HO basis 
increase to the nuclear response in the self-consistent relativistic random phase approximation (RRPA), which models the excited states by the superposition of particle-hole ($ph$) configurations interacting via the residual forces derived from the covariant energy density functional (CEDF)  \cite{RS.80,Ma2001}. The second subject is the nuclear response beyond RRPA, and the leading approximation here includes $ph\otimes phonon$ configurations, typical for the nuclear field theories \cite{BortignonBrogliaBesEtAl1977,BertschBortignonBroglia1983,KamerdzhievSpethTertychny2004,Tselyaev1989,LitvinovaTselyaev2007,Sol-book,Lenske:2019ubp,LoIudice2012}, in the relativistic framework, linking the dynamical effects of the induced interaction with CEDF and dubbed as relativistic time blocking approximation (RTBA) \cite{LitvinovaRingTselyaev2007}. Since the latter approach provides a better description of the nuclear resonances, it is important to quantify how the HO basis truncation sensitivity propagates from RRPA modes to complex configurations.

The second goal of our study is to define statistical errors and systematic
uncertainties in the description of nuclear response  stemming from the 
parameters of energy density functionals (EDFs) \cite{DNR.14,Stat-an,AAT.19}.
So far, systematic  errors and statistical uncertainties have been studied only for the ground 
state (such as binding energies, charge radii, two-neutron separation energies, 
neutron skins, and single-particle energies, etc.) and fission (such as potential energy curves)  
properties in non-relativistic Skyrme  \cite{Eet.12,GDKTT.13,NCNF.18,NCGNOT.20} and 
covariant DFTs  \cite{AARR.14,AAT.19,AARR.17}, but not for the nuclear response. We 
address this gap in understanding the EDF's error propagation in the second part of the 
manuscript, confining the first study of this kind to the RRPA response.

We chose a characteristic set of doubly magic nuclei across the nuclear chart from 
medium-light  $^{48,70}$Ca and $^{78}$Ni isotopes to the medium-heavy $^{132}$Sn 
nucleus and heavy $^{208}$Pb one, thereby including
moderately and extremely neutron-rich nuclear species, to assess the 
sensitivity of the results to the truncation of the HO basis in a broad range of the isospin asymmetry.
The response channels under study are the neutral resonances: the isoscalar 
$J^{\pi} = 0^+, 2^+$ and $3^-$, and the isovector $J^{\pi} = 1^-$ ones. 
These are the spectra associated with the parameters of the nuclear equation of state,
such as the nuclear compressibility, dipole polarizability, and symmetry 
energy, used to describe cosmic objects, particularly neutron stars.
Furthermore, the low-energy $1^-$ strength distributions in neutron-rich nuclei are responsible for the neutron capture rates during the r-process nucleosynthesis of heavy elements in stars, and the accurate information about the octupole collectivity and softness is critical for precision studies of the parity violation in the nuclear sector. These connections across the various fundamental physics frontiers further highlight the importance of placing accurate systematic and statistical uncertainties on the nuclear strength functions.

The paper is organized as follows. Section \ref{theory} concisely describes the theoretical 
framework used for nuclear response calculations. Section \ref{trunc-basis} encompasses 
the effects of the HO basis truncation on the single-particle states (Sec.\ \ref{trunc-basis-sp}) and
the RRPA (Sec.\ \ref{RRPA-part}) and RTBA (Sec.\ \ref{RTBA-part}) strength
functions.   Partial neutron and proton contributions to the RRPA response are 
analyzed in subsection \ref{partial}.
Statistical errors and systematic uncertainties in the calculated  
RRPA  strength functions
are analysed and compared in Section 
\ref{statistical-errors}. Finally, Section \ref{concl} summarizes the results
and sets the perspectives opened by this study. 

%%%%%%%%%%%%%%%%%%%%%%%%%%%%%%%%%%%%%%%%%%%%%
\section{Theoretical framework}
\label{theory}
%%%%%%%%%%%%%%%%%%%%%%%%%%%%%%%%%%%%%%%%%%%%%

Nuclear response theory is the optimal tool to describe nuclear spectral properties in a wide 
energy range. At the most basic level, it is confined to the random phase approximation 
(RPA) and its superfluid variant, quasiparticle RPA (QRPA). In the field-theoretical {\it ab initio} equation 
of motion (EOM) framework \cite{AdachiSchuck1989, DukelskyRoepkeSchuck1998} (Q)RPA is obtained 
by omitting the 
two-particle-two-hole ($2p2h$) and higher-rank correlations in the interaction kernel of the two-fermion 
propagator.  In the EOM of Rowe \cite{Rowe1968}, (Q)RPA is obtained by using the superposition of one-particle-one-hole  (two-quasiparticle) $1p1h$ (2q) excitation operators generating excited states by their action on a Hartree-Fock (Hartree-Fock-Bogoliubov) ground state.  (Q)RPA quantitatively reproduces the basic properties of giant resonances and soft modes, but for fine spectral details, higher complexity ($npnh$) correlations in both excited and ground states of the nucleus have to be included in the theory.

Such correlations are contained in the dynamical kernel of the {\it ab initio} EOM for the two-fermion response 
function and can be sorted and quantified based on systematic expansions.  Cluster decomposition of the 
operator strings comprising the dynamical kernel is one of the convenient possibilities \cite{LitvinovaSchuck2019,Litvinova2022}, which allows exact mapping to the quasiparticle-vibration coupling (qPVC). The latter comes with an emergent order parameter and allows, in principle, for perturbative expansion of fermionic self-energies. However, this is not free of unphysical singularities, so their non-perturbative treatment is preferred for numerical implementations \cite{Tselyaev1989,LitvinovaRingTselyaev2007,LitvinovaRingTselyaev2008, LitvinovaSchuck2019,Litvinova2022}.

The qPVC in the minimal coupling scheme includes $2q\otimes phonon$ configurations in the intermediate two-fermion propagator, representing the leading approximation beyond (Q)RPA. The vibrations, or phonons, are correlated $2q$ pairs, with the qPVC vertices as the order parameters.  This approach admits realistic implementations that employ effective interactions while keeping the algebraic structure of the {\it ab initio} theory. The effective interactions can be adjusted in the framework of the density functional theory or obtained from bare interactions. The use of effective interactions enables an efficient numerical algorithm based on the fact that reasonable phonons can be obtained already within (Q)RPA. These phonons are then used in the $2q\otimes phonon$ configurations, together with the mean-field quasiparticles and a subtraction, restoring the self-consistency of the {\it ab initio} framework \cite{Tselyaev2013}.

The first self-consistent microscopic approach, which includes (q)PVC in terms of $ph\otimes phonon$ ($2q\otimes phonon$) configurations, was presented in Refs. \cite{LitvinovaRingTselyaev2007,LitvinovaRingTselyaev2008} with applications to the monopole and dipole responses of medium-mass and heavy nuclei. These implementations were a major step toward a universal theory of nuclear structure rooted in particle physics, employing the effective meson-exchange interaction \cite{Lalazissis1997,NL3*}.  The approach was derived via the time blocking technique \cite{Tselyaev1989} applied to the phenomenological assumption about the leading role of $2q\otimes phonon$ configurations, and thus identified as relativistic (quasiparticle) time blocking approximation (R(Q)TBA). Later, both the phenomenological qPVC and time blocking were ruled out as unnecessary ingredients, as the complete response theory was obtained via {\it ab initio} EOMs \cite{LitvinovaSchuck2019,Litvinova2022}.  Furthermore, the latter relativistic EOM (REOM) developments enabled extensions to configurations of arbitrary complexity organized in (q)PVC.  The relativistic EOM confined by the $ph\otimes phonon$ ($2q\otimes phonon$) configurations REOM$^2$ with the (Q)RPA phonons is essentially equivalent to R(Q)TBA.  An example of such an extension was presented as REOM$^3$ accommodating $2q\otimes 2phonon$ configurations in Refs. \cite{LitvinovaSchuck2019,Litvinova2023a}.

In this work, focused on non-superfluid nuclei, we apply the REOM$^2$-RTBA version of the theory to the low-spin resonances. The results are to be compared to those obtained in RRPA. We consider the strength function defined as: 
\begin{eqnarray} 
S_F(\omega) &=& \sum\limits_{\nu>0} \Bigl[ |\langle \nu|F^{\dagger}|0\rangle |^2\delta(\omega-\omega_{\nu}) - |\langle \nu|F|0\rangle |^2\delta(\omega+\omega_{\nu})
\Bigr]
\nonumber \\
&=& -\frac{1}{\pi}\lim_{\Delta\to 0}\Im\sum\limits_{121'2'}F_{12}R_{12,1'2'}(\omega+i\Delta)F^{\ast}_{1'2'},
\label{SFF}
\end{eqnarray}
where the response function $R_{12,1'2'}(\omega)$ completely characterizes the nuclear structure. Its spectral representation reads:
\begin{eqnarray} 
R_{12,1'2'}(\omega) = \sum\limits_{\nu>0}\Bigl[ \frac{\rho^{\nu}_{21}\rho^{\nu\ast}_{2'1'}}{\omega - \omega_{\nu} + i\delta} -  \frac{\rho^{\nu\ast}_{12}\rho^{\nu}_{1'2'}}{\omega + \omega_{\nu} - i\delta}\Bigr],
\label{respspec}
\end{eqnarray} 
with the energies $\omega_{\nu} = E_{\nu} - E_0$ of the excited states $|\nu\rangle$ with respect to the ground 
state $|0\rangle$ energy and $\delta \to +0$. The matrix element $\langle \nu|F^{\dagger}|0\rangle$, for the typical 
one-body external field operator, 
\begin{eqnarray}
\langle \nu|F^{\dagger}|0\rangle &=& \sum\limits_{12}\langle \nu|F_{12}^{\ast}\psi^{\dagger}_2\psi_1|0\rangle = \sum\limits_{12}F_{12}^{\ast}\rho_{21}^{\nu\ast},\nonumber\\
\rho^{\nu}_{12} &=& \langle 0|\psi^{\dagger}_2\psi_1|\nu \rangle , 
\label{Frho}
\end{eqnarray}
is related to the transition densities $\rho^{\nu}_{12}$ expressed via the fermionic field operators $\psi_1$ and $\psi_1^{\dagger}$ in some single-particle basis, which is in our case the relativistic mean field basis.
Representing the Dirac delta-function by the Lorentz distribution
\begin{eqnarray}
\delta(\omega-\omega_{\nu}) = \frac{1}{\pi}\lim\limits_{\Delta \to 0}\frac{\Delta}{(\omega - \omega_{\nu})^2 + \Delta^2},
\end{eqnarray}
we get
\begin{eqnarray}
S(\omega) 
= -\frac{1}{\pi}\lim\limits_{\Delta \to 0} {\Im} \Pi(\omega+ i\Delta),
\label{SFDelta} \\
\Pi(\omega) 
=  \sum\limits_{\nu} \Bigl[ \frac{|\langle \nu|F^{\dagger}|0\rangle |^2}{\omega - \omega_{\nu}}
- \frac{|\langle \nu|F|0\rangle |^2_{\nu}}{\omega + \omega_{\nu} }
\Bigr].
\label{Polar}
\end{eqnarray}
In the numerical implementation, the finite smearing parameter $\Delta$ is used. Its value typically 
corresponds to the resolution of experimental data, which are of interest for comparison to the 
theory.  

The external field operators considered in this work are the isoscalar electric monopole ISE0 ($J^{\pi} = 0^+$), isovector (electromagnetic) electric dipole IVE1 ($J^{\pi} = 1^-$),  isoscalar electric quadrupole ISE2 ($J^{\pi} = 2^+$) and  isoscalar electric octupole ISE3 ($J^{\pi} = 3^-$) ones defined as:
\begin{eqnarray}
F^{(\text{ISE0})}_{00} &=& \sum\limits_{i=1}^A r_i^2Y_{00}({\hat{\bf r}}_i),\nonumber\\
F^{(\text{IVE1})}_{1M} &=& \frac{eN}{A}\sum\limits_{i=1}^Z r_iY_{1M}({\hat{\bf r}}_i) - \frac{eZ}{A}\sum\limits_{i=1}^N r_iY_{1M}({\hat{\bf r}}_i),
\nonumber\\
F^{(\text{ISEL})}_{LM} &=& \sum\limits_{i=1}^A r_i^LY_{LM}({\hat{\bf r}}_i), L \geq 2,
\label{Fext}
\end{eqnarray}
where $Z$ and $N$ are the numbers of protons and neutrons, respectively,  
$A = Z + N$,  $e$ is the proton charge  and $L$ is angular momentum.
These operators define the angular momentum and parity of the nuclear transitions, while the 
spectral features are determined by the response function (\ref{respspec}). Namely, the number 
of peaks in the resulting spectrum is associated with the number of terms in Eq. (\ref{respspec}), 
which, in turn, reflects the correlation content of the theory.  The positions of the peaks and 
transition densities, which define the peak heights, are thus found from the Bethe-Salpeter-Dyson 
equation (BSDE) for the response function, that is, in the operator form,
\begin{eqnarray}
R(\omega) = R^0(\omega) + R^0(\omega)\Bigl( K^0 + K^r(\omega) - K^r(0)\Bigr)R(\omega). \nonumber \\
\label{BSDE}
\end{eqnarray}
Here, $R^0(\omega)$ is the non-interacting particle-hole propagator determined by the mean field, and the term $K^r(0)$ eliminates the double counting of (q)PVC \cite{Tselyaev2013}.  The specific forms of the static and dynamic interaction kernels $K^0$ and $K^r(\omega)$ are given, for instance, in Refs. \cite{LitvinovaRingTselyaev2008,Litvinova2022}. For the cases discussed in this work, it is essential that the RRPA strength is obtained when neglecting the $K^r$ terms. The latter are included in RTBA in the $ph\otimes phonon$ approximation. In this work, we limit the description to such configurations 
 but note that a higher-complexity approach including superfluidity is available in Refs. \cite{LitvinovaSchuck2019,Litvinova2023a}.

   The RRPA and  RTBA calculations were performed with the NL3* meson-exchange 
interaction \cite{NL3*} in the relativistic mean field basis, 
defining the single-particle energies and associated Dirac spinors. Natural-parity RRPA phonons up to 
20 MeV with $J = [1,6]$ 
were collected to form the $ph\otimes phonon$ configurations in the $K^r(\omega)$ amplitude. 
The pure $ph$ configurations were included up to 100 MeV, while the $ph\otimes phonon$ ones 
were accommodated up to  45 MeV. 
To reveal the sensitivity of the strength functions of different multipolarities to the size of 
 the basis, the RRPA and  RTBA calculations are carried out with $N_F=20$ and $N_F=50$; 
the latter value being close to the maximum one achievable in the existing 
computer codes.

%%%%%%%%%%%%%%%%%%%%%%%%%%%%%%%%%%%%%%%%%%%%%
\section{The effects of the HO basis truncation on the calculated nuclear properties}
\label{trunc-basis}
%%%%%%%%%%%%%%%%%%%%%%%%%%%%%%%%%%%%%%%%%%%%%

%%%%%%%%%%%%%%%%%%%%%%%%%%%%%%%%%%%%%%%%%%%%%
\subsection{The energies of the single-particle states}
\label{trunc-basis-sp}
%%%%%%%%%%%%%%%%%%%%%%%%%%%%%%%%%%%%%%%%%%%%%

\begin{figure}[htb]
	\includegraphics[scale=0.23]{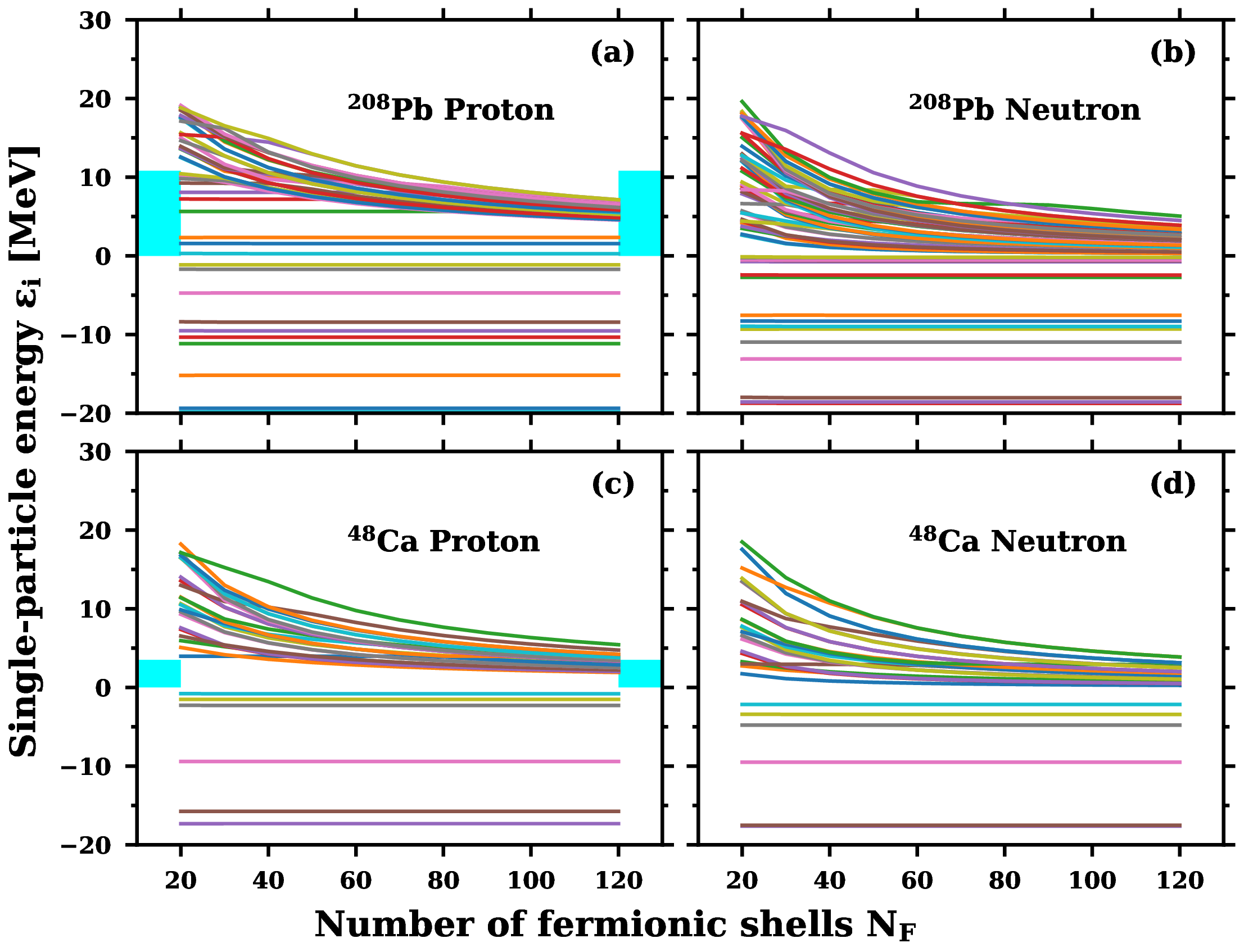}
 	\caption{The dependence of the single-particle energies of the proton 
	and neutron subsystems of $^{208}$Pb and $^{48}$Ca on the number of fermionic 
	shells $N_F$ employed in the calculations.  The calculations have been 
	carried out in steps of $\Delta N_F=10$. Note that for illustration purposes only  
	30 proton and 40 neutron positive energy single-particle 
	states, which are the lowest in energy at $N_F=20$	in the
	$^{208}$Pb, are shown. The number of positive energy states is reduced to 
	25 (the same for proton and neutron subsystems) in the case of $^{48}$Ca 
	nucleus.   	
	Color boxes in the left panels show the magnitude of the
	Coulomb barrier.}
\label{sp-energy-as-funct-NF-with-Coulomb}
\end{figure}

Fig.\ \ref{sp-energy-as-funct-NF-with-Coulomb} shows the evolution of the 
single-particle energies of the proton and neutron subsystems of the $^{208}$Pb and 
$^{48}$Ca nuclei with  increase of $N_F$. One can see that for $N_F\geq 20$, the single-particle 
energies of the bound states with negative energies almost do not change their values with 
increasing $N_F$. This is consistent with the fact that the calculated total binding energies of 
these nuclei either fully converge (the $^{48}$Ca nucleus) or are very close to a full convergence 
(the $^{208}$Pb nucleus) for $N_F=20$ \cite{NL5Z-DDMEZ-PCZ,OAD.25}. 

 However, the situation for positive energy states is different, and the details depend on 
the subsystem under consideration.  Let us start with the neutron subsystem.  One can see 
in Figs.\ \ref{sp-energy-as-funct-NF-with-Coulomb}(b) and (d) that with increasing $N_F$  the single-particle 
energies of the neutron states in $^{208}$Pb and $^{48}$Ca become lower, and that no full 
convergence of these energies is reached even at  $N_F=120$ for most of the states.  
Note that the increase of $N_F$ leads to the increase of the single-particle density above 
the zero energy threshold. 

%%%%%%%%%%%%%%%%%%%%%%%%%%%%%%%%%%%%%%%%%%%%%
%%%%%%%%%%%%%%%%%%%%%%%%%%%%%%%%%%%%%%%%%%%%%%%%

In contrast, the convergence  of positive energy proton single-particle 
states depend on whether they are located below or above the Coulomb 
barrier with the height $E_{Coul}$. With increasing $N_F$, the calculated proton 
single-particle states with  single-particle energies $e_i^{\pi}$ located below 
the Coulomb barrier (i.e. $0< e_i^{\pi} < E_{Coul}$) at $N_F=20$
typically change little or moderately,
see Figs.\ \ref{sp-energy-as-funct-NF-with-Coulomb}(a) and (c). This is because these
states are quasi-bound due to the presence of a finite Coulomb barrier.

 On the other hand,
the proton states located above the Coulomb barrier (i.e. $e_i^{\pi} > E_{Coul}$)
behave similarly to positive energy neutron states: with increasing $N_F$, their 
energies decrease, and the density of the single-particle states increases.
Remarkably, some proton single-particle states, which are located in the continuum
at $N_F=20$ become quasi-bound with increasing $N_F$ as one can also see in Figs.\ 
\ref{sp-energy-as-funct-NF-with-Coulomb}(a) and (c).
Note that the states located in the continuum experience larger changes in energy with
increasing $N_F$ as compared with those seen for the quasi-bound states.

  It is important to understand the origin of above discussed features of the 
behavior of the single-particle states with increasing $N_F$ for the neutron states
above zero energy threshold and proton states above the Coulomb barrier. It turns 
out that 
this behavior is due to the fact that the finite harmonic oscillator (HO) basis in nuclear 
many-body calculations effectively imposes a  hard-wall boundary conditions in the
coordinate space, i.e., it is equivalent to a spherical cavity of a radius $L_0$ 
\cite{FHP.12,BEHPW.16} 
\begin{eqnarray} 
L_0 = \sqrt{2(N_F+3/2)b}.  
\label{radius}
\end{eqnarray} 
in the case of spherical nuclei.  The radius of this cavity is defined by the oscillator frequency
$\hbar \omega_0$ and $N_F$ of the employed HO basis. Here, 
$b=\sqrt{\hbar/(m\omega_0)}$  is the oscillator length of the basis and $m$ denotes the 
nucleon mass. Note that Eq.\ (\ref{radius})  provides a rough estimate of $L_0$
(see Refs.\ \cite{MEFHP.13,OAD.25}).  
  
   It is interesting to compare the evolution of the energies of bound and unbound 
single-particle states seen in Fig.\ \ref{sp-energy-as-funct-NF-with-Coulomb} with 
that displayed in Fig.\ 9 of Ref.\ \cite{DNWBCD.96}. The latter figure shows their 
evolution  as a function of the radius $R_{box}$ of the spherical box employed in coordinate
space calculations. The comparison of  these two figures reveals significant similarities 
and suggests that the increase of $N_F$ in the HO basis, which leads to an increase
of the radius of the spherical cavity, is more or less equivalent to an increase of $R_{box}$ 
in coordinate space calculations.

   Note also that the recent analysis of neutron halos also points to the similarity of the description 
of such nuclei in the coordinate space and large HO bases calculations. For example,  the 
neutron densities of neutron halos in  $^{40}$Ne and  $^{72}$Ca obtained in coordinate space 
calculations of Refs.\ \cite{ZMR.03,PVLR.97}  are reproduced with high precisions in the HO 
calculations with $N_F=60$ and  $N_F=120$ bases, respectively (see Ref.\ 
\cite{DAO.25}). Spherical calculations in such HO  bases at the mean field level
are numerically cheap: they require 
only a few minutes of CPU time on a regular laptop. This is due to the moderate growth of the 
HO basis with the increase of  $N_F$ (see discussion of Table II in Ref.\ \cite{TOAPT.24}). 

%%%%%%%%%%%%%%%%%%%%%%%%%%%%%%%%%%%%%%%%%%%%%
\begin{figure}[htb]
	\includegraphics[scale=0.26]{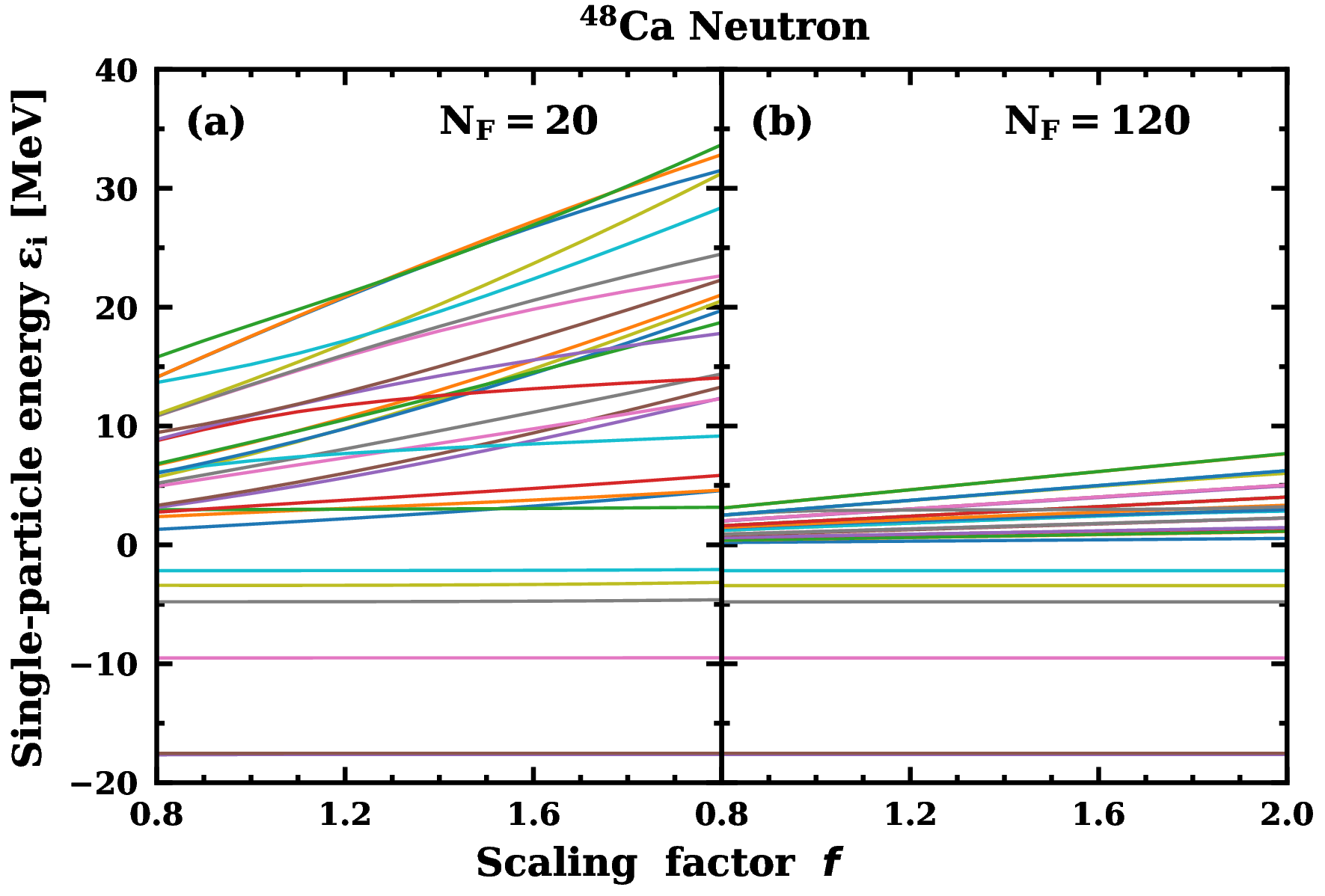}
 	\caption{The dependence of the energies of the neutron single-particle 
	states in $^{48}$Ca on the scaling factor $f$. The results of the calculations are 
	presented for $N_F=20$ and $N_F=120$. Note that the evolution of the energies of 
	only the 25 lowest at $f=1.00$ continuum states is displayed.}
\label{sp-energy-as-funct-of-f}
\end{figure}
%%%%%%%%%%%%%%%%%%%%%%%%%%%%%%%%%%%%%%%%%%%%%%%%

The oscillator frequency of the basis is defined by \cite{OAD.25}
\begin{eqnarray} 
\hbar \omega_0 = f \times 41 A^{-1/3} \qquad 
\end{eqnarray}    
where $f$ is the scaling factor. Thus, the radius $L_0$ of spherical cavity for large $N_F$
behaves as 
\begin{eqnarray}
L_0 \sim \sqrt{\frac{N_F}{\sqrt{f}}}
\end{eqnarray}
i.e., it increases with decreasing $f$. The numerical calculations with fixed 
$N_F$ but varying $f$ are presented in Fig.\  \ref{sp-energy-as-funct-of-f}.  The energies of  
bound single-particle states are almost not affected by the change of $f$. In contrast, the 
energies of continuum  states typically decrease more or less linearly  with decreasing $f$ 
leading to an increase in the density of the single-particle states at low $f$.
The increase of $N_F$ leads to a rise of the radius $L_0$ of the cavity and, as a consequence,
to a substantial suppression of the dependence of the energies of continuum states on $f$
[compare Figs.\ \ref{sp-energy-as-funct-of-f}(a) and (b)]. 

To bring the results of the HO basis calculations closer to those performed in coordinate space, 
it is desirable to increase the radius $L_0$ of the spherical cavity.  This can be done by either increasing 
$N_F$ or by decreasing $f$ or by the combination of both.  However, too low values of $f$ (i.e. $f <1.0$)
lead to numerical instabilities at the mean field level \cite{OAD.25}. Thus, the value of $f=1.0$ is
used in the present calculations of nuclear response. Such a value was used in relativistic calculations 
of the ground state properties and nuclear response for more than thirty years. Note that a higher value of 
$f \sim 1.4$ is recommended for the calculations of binding energies based on recent global optimization
of the HO basis \cite{OAD.25}. This reflects the different sensitivity of the bound and positive energy 
single-particle states to the properties of the HO basis: while the convergence of the former is defined 
by nucleonic potential, that of the latter depends on hard-wall boundary conditions in the coordinate space 
(spherical cavity), effectively imposed by the HO  basis in nuclear many-body calculations.

%%%%%%%%%%%%%%%%%%%%%%%%%%%%%%%%%%%%%%%%%%%%%
\subsection{Nuclear response: RRPA}
\label{RRPA-part}
%%%%%%%%%%%%%%%%%%%%%%%%%%%%%%%%%%%%%%%%%%%%%

   The impact of the dependence of the positive energy single-particle states on 
the size of the fermionic basis affecting physical observables of 
interest has not been investigated so far in nuclear response calculations.
 It influences the continuum effects in the calculations formulated
in a basis set expansion formalism.
Alternatively, the continuum can be partly included in the coordinate
space representation: this is done in Refs.  \cite{Shlomo1975,Kamerdzhiev1998,Matsuo2002} 
and references therein via numerical implementations with free and Coulomb asymptotics 
of the single-nucleon wave functions.
 It was established  in these calculations that the presence of configurations with a 
 nucleon in the continuum adds a natural envelope to the excited states above the particle emission 
 threshold, which is relatively small (100-200 keV) in medium-mass and heavy nuclei and sensibly larger 
 (a few MeV) in light nuclei. However, only partial success in the description of the giant resonances was achieved with this treatment of the continuum, even after adding complex configurations \cite{LitvinovaTselyaev2007}. 
This is because it is necessary to include the multi-particle 
continuum in the theory due to the opening of multi-particle escape channels in the energy 
regime of broad giant resonances. However, this is a very challenging task, which has not
been done so far.
 As we show in the following, direct calculations in extended 
bases represent an attractive alternative 
to the coordinate space calculations since
the continuum effects can be automatically included on (Q)RPA and (Q)RPA+(q)PVC levels. 
Moreover, a significant part of the multi-particle continuum can be potentially taken into account in
such calculations in the HO basis in a simpler way than in the coordinate space representation.

We have chosen the natural-parity low-spin excitations for this first study, namely the 
ISE0,  IVE1, ISE2, and  ISE3 channels of response associated with the excitation operators (\ref{Fext}). 
The typical excitation spectra for each of the natural spin-parity channels contain a high-frequency 
giant oscillation and a low-frequency soft mode; however, these modes have different relative 
positions and strengths in different channels \cite{harakeh2001giant,RS.80}. 
First, we concentrate on the most basic relativistic approach, RRPA, which represents 
the major building block for more advanced approaches. In RPA theories, the response is 
comprised of coherent particle-hole transitions, and the structure of the resonances is 
dominated by the Landau damping \cite{RS.80}.

  Five doubly magic nuclei were chosen for this study: $^{48,70}$Ca, $^{78}$Ni, 
$^{132}$Sn, and  $^{208}$Pb nuclei.  They differ substantially in the $N/Z$ ratio
ranging from 1.4 for $^{48}$Ca to 2.5 for $^{70}$Ca, in the position of neutron 
chemical potential with respect of continuum threshold and in the position of
proton chemical potential with respect of the saddle point of the Coulomb barrier
(see  Table \ref{Table-sel}). This selection allows the study of the mass and neutron 
excess dependence of basis truncation errors in nuclear response calculations.

%%%%%%%%%%%%%%%%%%%%%%%%%%%%%%%%%%%%%%%%%%
\begin{table}
\caption{
The values of the $N/Z$ ratio, neutron ($\lambda_n$) and proton ($\lambda_p$) chemical potentials and the heights of the Coulomb barrier $V_{coul}$ in 
indicated nuclei. Note that since pairing is neglected in the present calculations, the $\lambda_i$ 
value is equal to the energy of the last occupied single-particle  state in the $i$-th 
subsystem of the ground state. In addition, column 6  shows the 
 $V_{coul}-\lambda_p$ for the proton subsystem.}
 \label{Table-sel}
 \vspace{0.5cm}
	\begin{tabular}{ c c c c c c }
	\hline 
	Nucleus & N/Z & $\lambda_n$ & $\lambda_p$ & $V_{coul}$ & $V_{coul}-\lambda_p$ \\ 
	\hline 
	    1        &  2     &   3             &        4             &         5  & 6 \\          
	$^{40}$Ca  & 1.0 & -16.624 & -9.026 & 3.683 & 12.709 \\ 
	\hline 
	$^{48}$Ca  & 1.4 & -9.641 & -15.549 & 3.526 & 19.075 \\ 
	\hline 
	$^{70}$Ca  & 2.5 & -0.565 & -29.374 & 2.882 & 32.256 \\ 
	\hline 
	$^{78}$Ni  & 1.79 & -5.767 & -21.525 & 4.406 & 25.931 \\ 
	\hline 
	$^{132}$Sn & 1.64 & -7.847 & -16.084 & 7.230 & 23.314 \\ 
	\hline 
	$^{208}$Pb & 1.54 & -7.723 & -7.761 & 10.843 & 18.604 \\ 
	\hline 
	\end{tabular} 
\end{table}
%%%%%%%%%%%%%%%%%%%%%%%%%%%%%%%%%

%%%%%%%%%%%%%%%%%%%%%%%%%%%%%%%%%%%%%%%
\begin{figure}[htb]
	\includegraphics[scale=0.45]{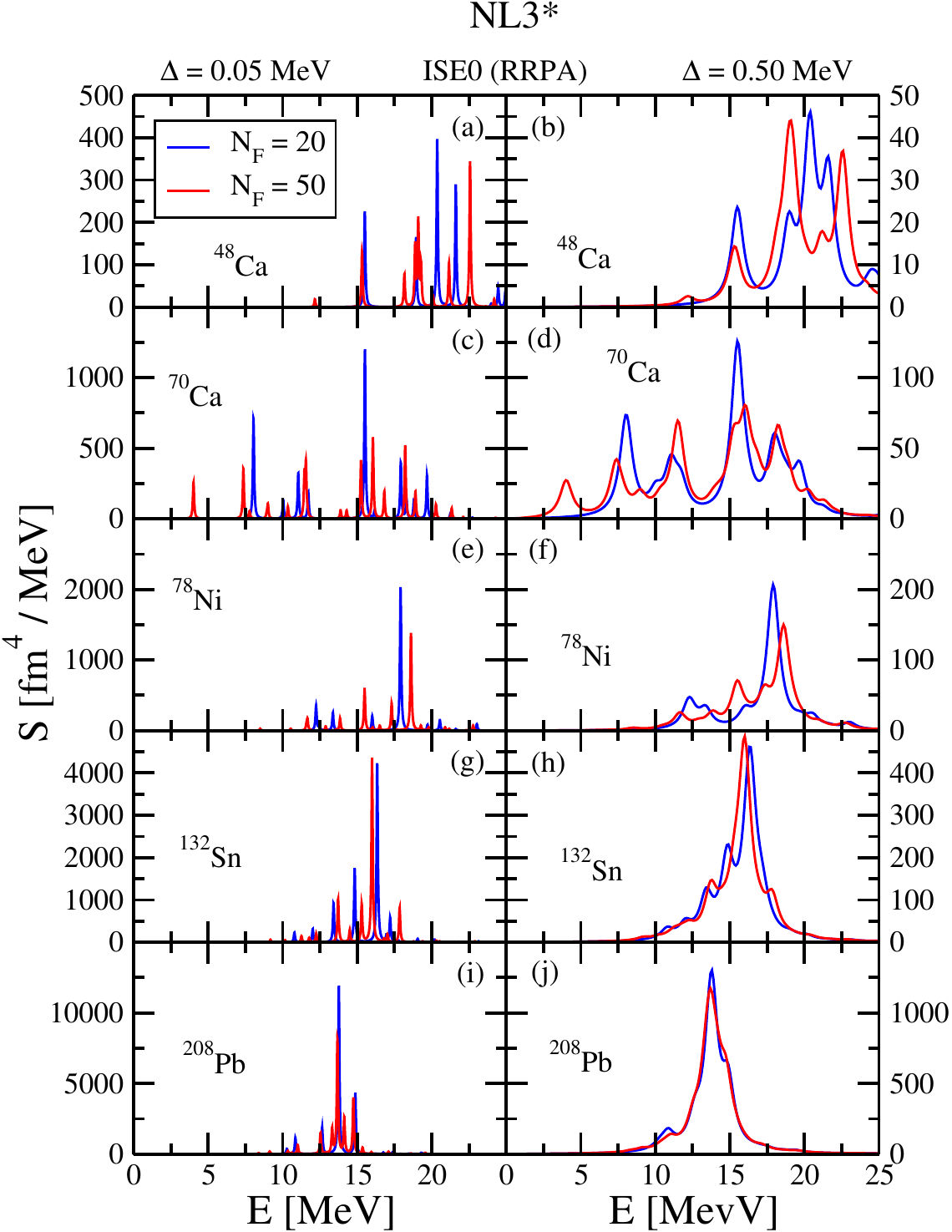}
 	\caption{ISE0 strength functions $S(E)$ obtained in RRPA
	 for the indicated nuclei. The results obtained with $N_F=20$ and $N_F=50$
	 are shown by blue and red curves, respectively. Left and right panels
	 show the results with $\Delta = 0.05$ and $\Delta =0.5$ MeV, respectively.}
\label{Strength-ISGMR}
\end{figure}
%%%%%%%%%%%%%%%%%%%%%%%%%%%%%%%%%%%%%%%%%%%%%%%%

The ISE0 strength functions obtained in the RRPA calculations with two values of the smearing  parameter, 
$\Delta$ = 50 keV and  $\Delta$ = 500 keV, characterizing the fine and gross details of each 
distribution, are presented in Fig. \ref{Strength-ISGMR}.  The high-energy part of the E0 response is 
the isoscalar giant monopole resonance (ISGMR), which corresponds to a uniform radial oscillation 
of the nucleus. This is the famous breathing mode quantifying nuclear compressibility. In this resonance, 
all nucleons move coherently in phase, producing a ``breathing mode'' in which the nuclear radius 
expands and contracts while preserving spherical symmetry. This mode is often probed by hadronic 
scattering at forward angles  \cite{GARG201855}.

One can see from Figs. \ref{Strength-ISGMR} (b), (d), (f), (h), and (j) that the gross structure of the ISGMR 
is significantly affected by the HO basis size. In particular, in $^{48}$Ca, the strength 
at $E\approx 20$ MeV becomes redistributed  between the  two major peaks, and a soft mode appears 
at $E\approx 12$ MeV in the calculations with $N_F=50$.
In $^{70}$Ca with  an extreme ratio of $N/Z=2.5$, these effects are further enhanced: the high-energy 
strength redistribution is more pronounced, and the appearance of the low-energy mode is more evident. 
The situation  further evolves when eight protons are added to $^{70}$Ca, leading to $^{78}$Ni,  which 
has  the same number of neutrons (i.e., $N=50$) as $^{70}$Ca:  a single peak structure 
of ISGMR in the $N_F=20$ calculations is replaced by a fragmented double peak structure when 
$N_F=50$  is used, and a much weaker soft low-energy mode appears in the calculations with $N_F=50$
[see Fig.\  \ref{Strength-ISGMR} (f)]. 
This can be attributed to the relevance of the asymmetry parameter $I = (N-Z)/A$, which is 
directly associated with the involvement of loosely bound neutron orbitals in the formation of the excited 
states.  The differences between the strength functions calculated with $N_F=20$ and $N_F=50$ decrease 
with increasing mass. In $^{132}$Sn,  the peak of the ISGMR moves down in energy by
approximately 1 MeV in the calculations with $N_F=50$ but the shapes of the strength functions are
comparable in the $N_F=20$ and $N_F=50$ results [see Fig. \ref{Strength-ISGMR}(h)].
  
In $^{208}$Pb, the difference between the strength functions obtained with $N_F=20$ and $N_F=50$ is 
rather small.   The left column of Fig. \ref{Strength-ISGMR} [panels (a),(c),(e),(g) and (i)] displays refined 
ISE0 spectra of the same nuclei to reveal the mechanism of the strength redistribution when $N_F$ increases
from 20 to 50. With such a modification, the energies of the excited states change, and more excited states 
appear in the energy window $0 \leq E \leq 25$ MeV, which causes the change in the pattern of the fragmentation 
of the resonances. Again, this effect is more pronounced in light nuclei, where the ISGMR centroid also changes, reflecting some influence of continuum states on nuclear compressibility.

%%%%%%%%%%%%%%%%%%%%%%%%%%%%%%%%%%%%%%%%%%%%%
\begin{figure}[htb]
	\includegraphics[scale=0.45]{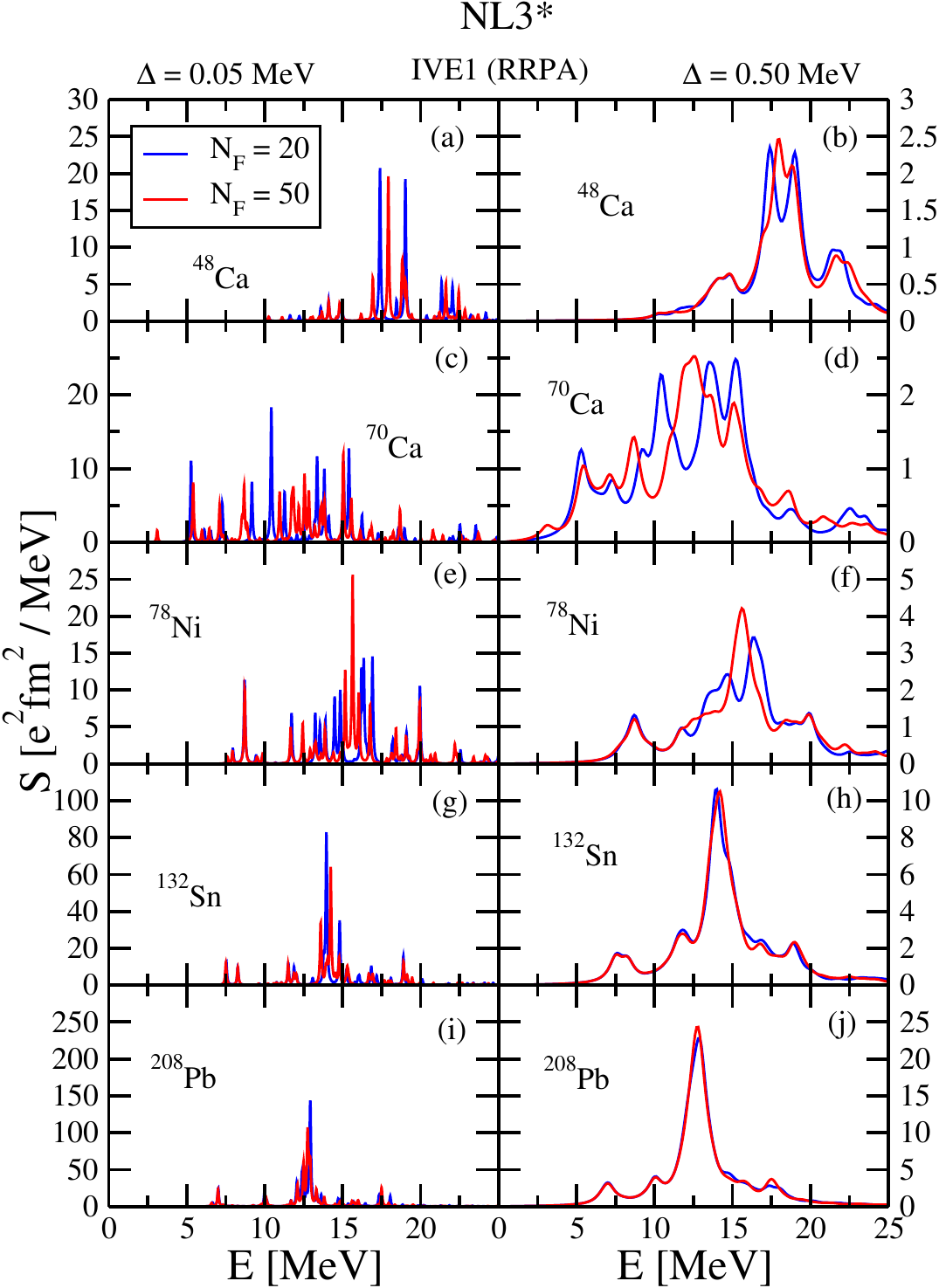}
 	\caption{
The same as in Fig.\ \ref{Strength-ISGMR}  but for
	IVE1.}
\label{Strength-IVGDR}
\end{figure}
%%%%%%%%%%%%%%%%%%%%%%%%%%%%%%%%%%%%%%%%%%%%%%%%

The RRPA IVE1 spectra of the same nuclei are presented in Fig. \ref{Strength-IVGDR}, organized similarly to the ISE0 ones.  The high-energy part of the IVE1 response is known as the isovector giant dipole resonance (IVGDR),
protons and neutrons oscillate out of phase, generating a time-dependent electric dipole moment. In experiments, this mode is most sensitively probed by photo-absorption and Coulomb excitation.
The IVGDR, its low-energy counterpart pygmy dipole resonance (PDR), and related observables, such as the dipole polarizability, constrain the symmetry energy and neutron skin thickness \cite{piekarewicz2011pygmy,Roca2015,fattoyev2018neutron}.

   The overall observations which follow from the analysis of Fig.\ \ref{Strength-IVGDR}  is that 
on average the IVE1 excitations are less affected by the HO basis increase 
than the ISE0 ones.  The IVE1 strength is almost not affected by the change of 
$N_F$ from 20 to 50 in $^{132}$Sn and, especially, in $^{208}$Pb [see Figs.\ 
\ref{Strength-IVGDR}(h) and (j)]. This change of basis triggers smaller changes of the 
IVE1 strength functions as compared with the ISE0 ones in $^{48}$Ca and $^{78}$Ni  
[compare Figs.\ \ref{Strength-IVGDR}(b) and (f) with Figs.\ \ref{Strength-IVGDR}(b) and (f)].
However, in these nuclei it leads to somewhat larger fragmentation of the IVE1 strength functions
in the $N_F=50$ results as compared with the $N_F=20$ ones. It is only in the $^{70}$Ca 
nucleus that the changes of the IVE1 strength function caused by the increase of basis
size are substantial. They are due to the fact that the neutron Fermi level is located close 
to the neutron continuum (see Table \ref{Table-sel}). As a consequence, the absolute majority of the 
$ph$ excitations that build the strength function involve the excitations to the neutron continuum 
(see Sec.\ \ref{partial} below) and the modifications in it caused by the increase of $N_F$ have 
a substantial impact on the strength function.

%%%%%%%%%%%%%%%%%%%%%%%%%%%%%%%%%%%%%%%%%%%%%
\begin{figure}[htb]
	\includegraphics[scale=0.44]{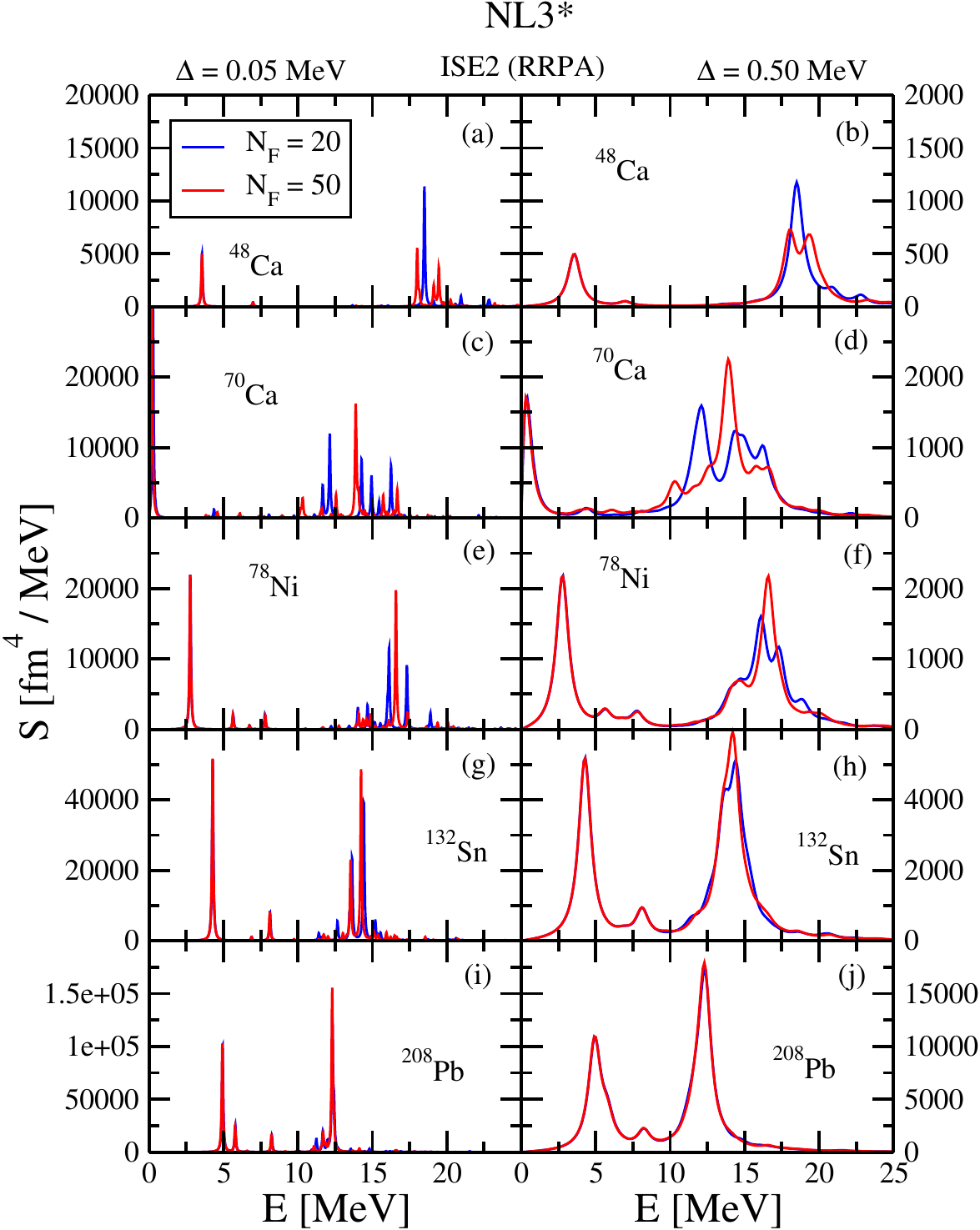}
 	\caption{
The same as in Fig.\ \ref{Strength-ISGMR}  but for
	ISE2.}
	\label{Strength-ISGQR}
\end{figure}
%%%%%%%%%%%%%%%%%%%%%%%%%%%%%%%%%%%%%%%%%%%%%%%%

The ISE2 strength distributions obtained in the RRPA calculations are displayed in Fig. \ref{Strength-ISGQR} 
for the same nuclei and the same two values of the smearing parameter. The isoscalar quadrupole 
response represents the cleanest case of the generic picture of two-peak structure, where both the 
high-energy and low-energy peaks are strongly collective. The latter appears around 
$0\hbar\omega$ and often resolves a dominant state, while the latter one is grouped around
$2\hbar\omega$. The $0\hbar\omega$ low-energy $2^+$ states are formed predominantly by the 
superposition of nucleonic transitions within the same shell in open-shell and heavy nuclei or between 
the neighboring shells of the same parity, for instance, between the occupied $\nu 1f_{7/2}$ state 
forming itself a shell and the neighboring $fp$ shell in $^{48}$Ca. The $2\hbar\omega$ branch is the 
isoscalar giant quadrupole resonance ISGQR,  formed by coherent transitions across a major oscillator 
shell. Macroscopically, both the $0\hbar\omega$ and $2\hbar\omega$ branches involve collective shape 
oscillations, when the nucleus oscillates between prolate and oblate deformations, conserving volume 
while changing shape. The ISGQR is typically excited by inelastic $\alpha$ scattering and electron scattering.

The lowest collective $2^+_1$ states are placed between 2.5 and 5 MeV in our group of nuclei, except for 
$^{70}$Ca  [see Fig. \ref{Strength-ISGQR}]. As RRPA is not a complete theory, the $2^+_1$ 
energies are typically higher than the observed ones, but are rather reasonable. The placement of the 
$2^+_1$ state in $^{70}$Ca near zero energy indicates substantial quadrupole softness, which is confirmed 
by the analysis of potential energy surfaces in axial  relativistic Hartree-Bogoluibov calculations.
The effect of the $N_F$ increase is almost non-existent for the lowest $2^+_1$ states 
in all nuclei, minimal on the ISGQR in $^{132}$Sn and extremely small for the ISGQR in $^{208}$Pb.
In contrast, the increase of $N_F$ visibly increases the fragmentation of ISGQR in the $^{48}$Ca 
nucleus but  suppresses it in very neutron-rich $^{70}$Ca and $^{78}$Ni nuclei.

%%%%%%%%%%%%%%%%%%%%%%%%%%%%%%%%%%%%%%%%%%%%%
\begin{figure}[htb]
	\includegraphics[scale=0.42]{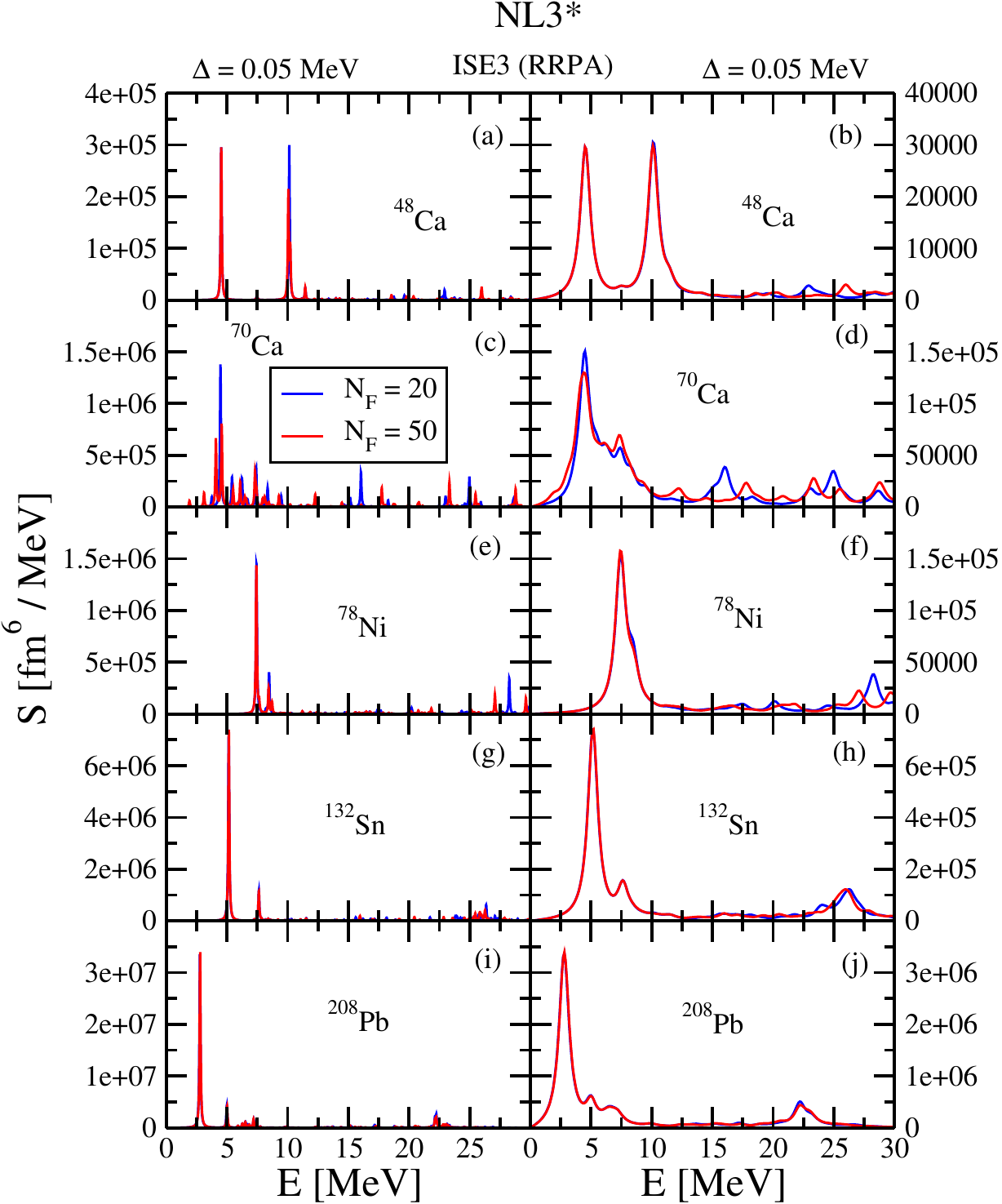}
 	\caption{
The same as in Fig.\ \ref{Strength-ISGMR}  but for
	ISE3.}
	\label{Strength-octupole}
\end{figure}
%%%%%%%%%%%%%%%%%%%%%%%%%%%%%%%%%%%%%%%%%%%%%%%%

The ISE3 or octupole strength is displayed in Fig. \ref{Strength-octupole}.
 Similarly to the quadrupole response, the octupole one splits into two main concentrations of strength in the $1\hbar\omega$ and $3\hbar\omega$ regions. 
 Both of them are groups of states; however, the lower-energy part of ISE3 is dominant in contrast to the lower-spin resonances. The high-frequency ISGOR is only nominally giant regarding its intensity
since the major part of the octupole strength is located below 10 MeV. The response to the octupole operator corresponds to higher-order surface vibrations in which 
the nucleus undergoes reflection-asymmetric, pear-shaped oscillations with protons and neutrons moving in phase. The ISE3 is mainly excited by inelastic scattering of isoscalar probes sensitive to surface properties and higher-order collectivity in nuclear dynamics.

 One can see from Fig. \ref{Strength-octupole} that the major, i.e., low-energy,  ISE3 response is not affected 
by the continuum and by the change of the basis size, except for the case of $^{70}$Ca.
 This is similar to the case of ISGQR. 
With the exception  of $^{132}$Sn and  $^{208}$Pb, the fragmentation pattern of the 
ISGOR is affected considerably 
when $N_F=50$ is used instead of $N_F=20$.

%%%%%%%%%%%%%%%%%%%%%%%%%%%%%%%%%%%%%%%%%%%%%
\subsection{Nuclear response: RTBA}
\label{RTBA-part}
%%%%%%%%%%%%%%%%%%%%%%%%%%%%%%%%%%%%%%%%%%%%%

As mentioned above, RRPA is not complete but rather a basic response theory; however, it is used as a 
major  building block for extended approaches, which include correlations beyond RRPA. The approximations 
beyond RRPA use the RRPA solutions, which are interpreted as basis vibrational states (phonons) to form 
complex configurations beyond the particle-hole ones. These approaches constitute more realistic 
descriptions of the nuclear response, and therefore, it is interesting to see how the effects of the modification 
of the continuum single-particle states caused by the increase of the size of the basis propagate 
to complex configurations.

%%%%%%%%%%%%%%%%%%%%%%%%%%%%%%%%%%%%%%%%%%%%%
\begin{figure}[htb]
	\includegraphics[scale=0.35]{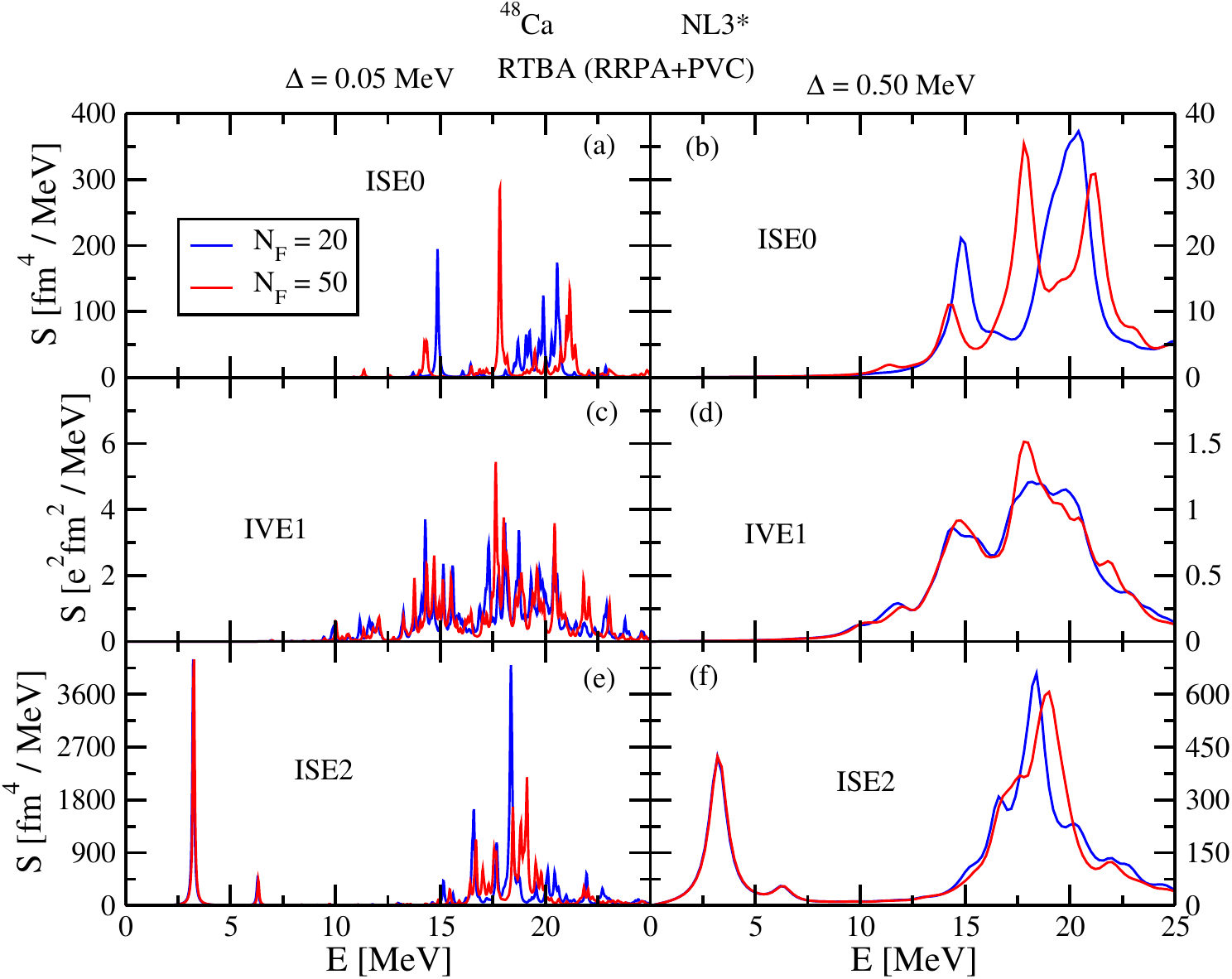}
 	\caption{
The dependence of the RTBA results on the size of fermionic basis
	in $^{48}$Ca. The strength functions of the ISE0, IVE1, and ISE2 resonances 
	calculated with 20 and  50 harmonic oscillator shells are displayed. The left 
	panels show the strength functions obtained with $\Delta$ = 50 keV, and the 
	right panels show those with $\Delta$ = 500 keV.	
	\label{RTBA-48Ca}
}
\end{figure}
%%%%%%%%%%%%%%%%%%%%%%%%%%%%%%%%%%%%%%%%%%%%%%%%

%%%%%%%%%%%%%%%%%%%%%%%%%%%%%%%%%%%%%%%%%%%%%
\begin{figure}[htb]
	\includegraphics[scale=0.35]{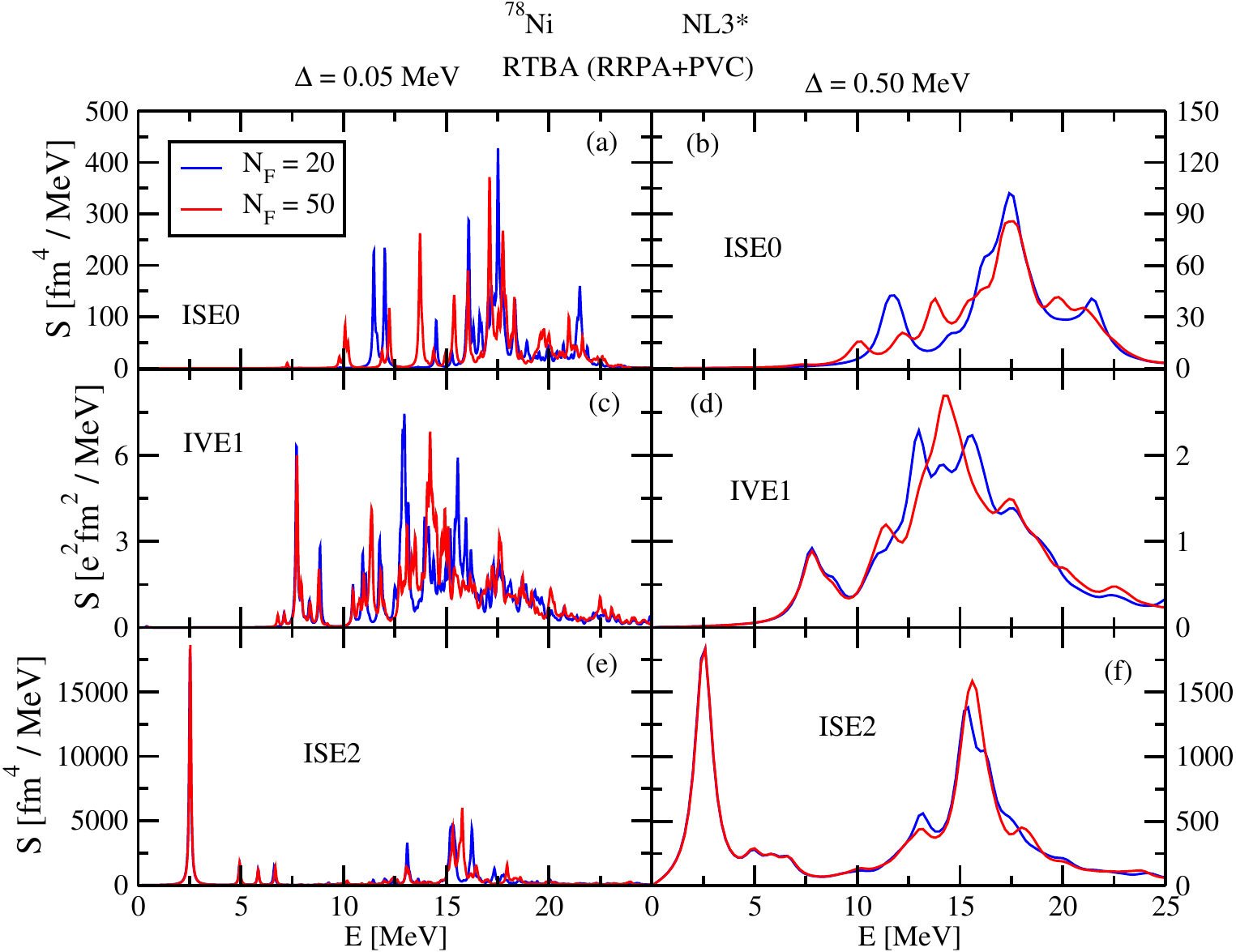}
 	\caption{
The same as in Fig.\ \ref{RTBA-48Ca} but for $^{78}$Ni.
	\label{RTBA-78Ni}
}
\end{figure}
%%%%%%%%%%%%%%%%%%%%%%%%%%%%%%%%%%%%%%%%%%%%%%%%

%%%%%%%%%%%%%%%%%%%%%%%%%%%%%%%%%%%%%%%%%%%%%
\begin{figure}[htb]
	\includegraphics[scale=0.35]{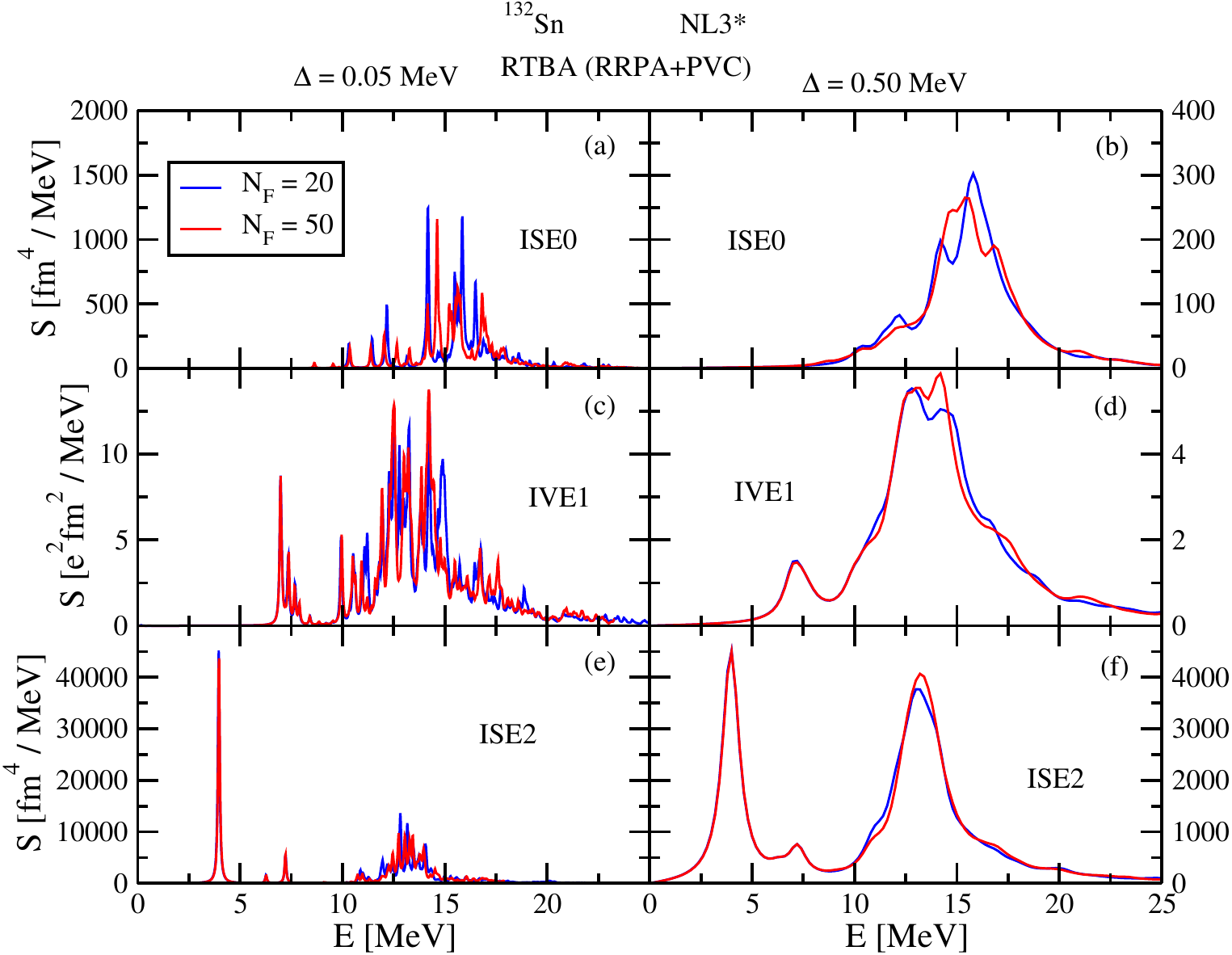}
 	\caption{
The same as in Fig.\ \ref{RTBA-48Ca} but for $^{132}$Sn.
	\label{RTBA-132Sn}
}
\end{figure}
%%%%%%%%%%%%%%%%%%%%%%%%%%%%%%%%%%%%%%%%%%%%%%%%

RTBA calculations with two values of the smearing parameter, $\Delta$ = 50 keV and 
$\Delta$ = 500 keV, are presented in Figs. \ref{RTBA-48Ca}, \ref{RTBA-78Ni}, and \ref{RTBA-132Sn} for the 
isoscalar monopole,  isovector dipole, and isoscalar quadrupole  
resonances in $^{48}$Ca, $^{78}$Ni, and $^{132}$Sn, respectively.
We compare the strength 
functions obtained with $N_F = 20$ and $N_F = 50$ fermionic harmonic oscillator shells, 
particularly to reveal the influence of the 
HO basis size associated with 
the continuum influence on gross and fine features of the nuclear resonances calculated beyond RRPA.

The difference first appears in the density and completeness of the RRPA phonon spectra 
of all multipolarities, which increase with the rise in the number of shells to $N_F = 50$. 
This is  consistent with the observation that in the RRPA calculations, the low-energy 
states remain intact, but the high-energy ones redistribute and densify. Because of that, in 
addition to the appearance of more particle states in the relevant energy window, the dynamical 
kernel  acquires a larger number of $ph\otimes phonon$ configurations on the high-energy side. 
The second effect is a visible rearrangement of the strength distribution in the main channel, which 
again originates from the RRPA primordial states before they undergo fragmentation. The resulting 
spectra show an interplay of these two effects.

The most pronounced influence of the $N_F$ increase is seen on the resonances in $^{48}$Ca 
shown in Fig. \ref{RTBA-48Ca}. The case of ISE0 in $^{48}$Ca is of special interest because it 
represents a challenge for the unified description of nuclear compressibility \cite{Li2022,Litvinova2023}. 
Here, one can see a major effect of the $N_F$ increase, in particular, on the ISGMR's centroid placement.
The description of the ISE0 centroid in a self-consistent way across the nuclear chart is of prime importance 
for the characterization of nuclear compressibility and equation of state.  Here, we just note the sensitivity 
of the ISE0 centroid in light nuclei to the oscillator basis completeness and continuum effects, while a more 
focused study of monopole spectra will be done in a dedicated work. Overall,  both the gross features and 
the fragmentation pattern of the ISE0 response of $^{48}$Ca change considerably with extending the 
$N_F$ value.
  Some softening and fragmentation of the low-energy monopole peak occurs, but it remains identifiable. 
The dipole and quadrupole responses of $^{48}$Ca also show sensitivity to the $N_F$ value. 
Its impact on the gross feature of these spectra is less visible; however, the fragmentation patterns of the 
IVE1 and ISE2 become richer, which is a direct consequence of the growing number of $ph\otimes phonon$ configurations.

The rearrangement of the low-lying E1 states is of special interest as it can be measured with high resolution \cite{DeryaSavranEndresEtAl2014}. Higher configuration complexity 
calculations, including $ph\otimes 2phonon$ configurations with $N_F$ = 20, were presented in Ref. \cite{LitvinovaSchuck2019} and showed the importance of such configurations for an accurate description of the low-energy E1 strength in $^{48}$Ca as well as some potential for an improved numerical implementation. Here, we see that the HO basis completeness plays an important role in the redistribution of the PDR, so that  the $ph\otimes 2phonon$ approach will likely be sensitive to $N_F$ as well.

The results for $^{78}$Ni are presented in Fig. \ref{RTBA-78Ni}. This nucleus has a considerably 
larger isospin asymmetry, so that one expects an enhancement of the 
effects associated with the increase of $N_F$.
Indeed,  the formation of a pronounced soft monopole mode is seen in this nucleus and the increase of $N_F$
leads to its considerable softening and fragmentation.
In the $^{48}$Ca and  $^{78}$Ni nuclei, there is a physical possibility of the formation 
of measurable breathing-type neutron skin oscillations.
Such measurements are definitely more difficult in $^{78}$Ni, but, in principle, still are possible
with radioactive beams in inverse kinematics. They, in combination with theory, would help to
extrapolate the quantification of nuclear compressibility to asymmetric nuclear matter. 
The ISGMR  also receives additional fragmentation with the increase of $N_F$. The IVE1 spectrum is of 
interest in the context of the r-process nucleosynthesis, in which $^{78}$Ni represents a typical waiting point.
The pygmy dipole resonance is almost unaffected by the continuum states, while the ISGDR acquires a smoother envelope, washing out the structure artifacts still present at $N_F = 20$. Interestingly, a similar effect is seen in the E2 channel, pointing to the important role of the continuum states in the formation of the ISGQR.

Fig. \ref{RTBA-132Sn} displays the results of  analogous calculations for $^{132}$Sn which is another 
neutron-rich nucleus known as an r-process waiting-point. As it is 
much heavier, 
a smaller influence of the continuum states is expected.  Indeed, 
the effect of the $N_F$ increase is minimal on the dipole and quadrupole excitation spectra, but the 
ISE0 strength still shows some sensitivity to it. One can notice a stronger fragmentation of the giant 
monopole resonance and the appearance of the low-energy states between 5 and 10 MeV caused 
solely by the $N_F$ increase.
 The latter represents a measurable effect and can be tested in high-resolution measurements when they are available for this nucleus. 
 The RTBA calculations for the monopole response of $^{132}$Sn are especially interesting in the context of the accurate placement of the ISGMR's centroid. In particular, the study of Ref. \cite{Litvinova2023} suggested a direct correlation between the collectivity of the  $2^+_1$ states and the shift of the ISGMR's centroid due to qPVC, which are most pronounced in mid-shell nuclei. In this context, the strength distributions shown in Figs.  \ref{RTBA-132Sn} (a) and (b) are in agreement with the general trend. The qPVC effect is visible when comparing these strength distributions with those in Fig. \ref{Strength-ISGMR} and shows no considerable shift of the ISGMR's centroid, which is consistent with the absence of the quadrupole softness in $^{132}$Sn. Further, the HO basis increase produces an interesting enrichment of the fine structure of the ISGMR but does not affect the centroid placement. The dipole and quadrupole spectra are minimally 
affected within the approach confined by $ph\otimes phonon$ configurations. 
The experimental data on the E1 response are available in Ref. \cite{Adrich2005}, and they are in general agreement with the RTBA calculations of Ref. \cite{LitvinovaRingTselyaev2007}. The statistics of these measurements, however, were limited and, thus, do not allow for firm conclusions regarding the fine structures of PDR and IVGDR, which thereby remains a task for the future. On the theory side, including higher
configuration complexities would be desirable to evaluate the role of the multi-particle continuum in both dipole and quadrupole channels of response.

The obtained effects were qualitatively expected as the spectra of light nuclei are known to be sensitive to the continuum and, as a result, have sensible escape widths. Describing the continuum effects on giant resonances and soft modes in the response theory frameworks was attempted in the past, but these studies were confined by the single-particle continuum, see, for instance, \cite{Shlomo1975,Kamerdzhiev1998}. The direct RTBA computation with a large number of harmonic oscillator shells relaxes this limitation, as all the continuum single-particle states are included in the complex configurations of the dynamical kernel. In this context, the present study paves the way for investigating many-body continuum effects on the nuclear resonances.
A desirable addition to the present machinery would be accurate free asymptotics of the continuum states used in Refs. \cite{Shlomo1975,Kamerdzhiev1998,Daoutidis2009,DaoutidisRing2011}, which would enable an assessment of the limitations of employing the HO basis.

%%%%%%%%%%%%%%%%%%%%%%%%%%%%%%%%%%%%%%%%%%%%%%%%
\subsection{Isospin asymmetry and partial neutron and proton contributions: RRPA response}
\label{partial}
%%%%%%%%%%%%%%%%%%%%%%%%%%%%%%%%%%%%%%%%%%%%%%%%

\begin{figure}[htb]
	\includegraphics[scale=0.35]{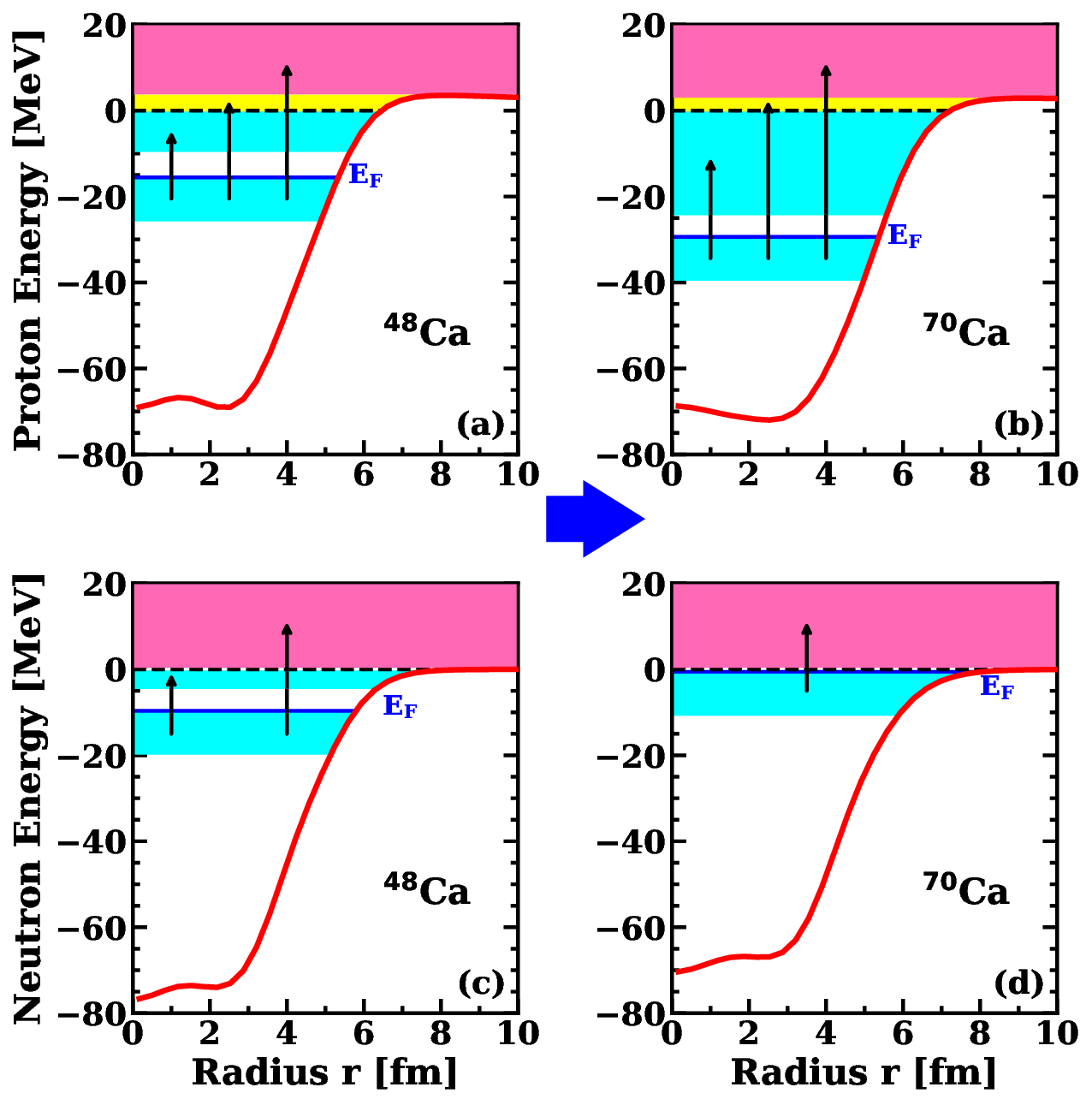}
	\caption{Schematic illustration of proton and neutron particle-hole
	excitations forming excitation spectra in indicated nuclei and their 
	evolution with neutron number. Proton and neutron potentials and 
	the positions of respective Fermi levels (see Table \ref{Table-sel}) are 
	obtained in the RMF calculations without pairing.	
	The energy ranges of the single-particle
	states from which (and to which) particles can be excited are 
	schematically shown by shaded areas. Cyan, yellow and pink
	colors are used for particle bound, particle quasi-bound and
	continuum states.  The $ph$ excitations are schematically  shown
	by arrows.
\label{fig-schematic}
}
\end{figure}

    It is important to understand above discussed features of the impact of the
HO basis increase on the strength functions $S(E)$ obtained in the RRPA and
RTBA calculations and their evolution with neutron excess and mass.
To do that the schematic illustration of the modifications of proton and neutron $ph$ 
excitations involved in building the nuclear response with increasing neutron number 
from $N=28$ to $N=50$ is shown in Fig.\  \ref{fig-schematic}.   

%%%%%%%%%%%%%%%%%%%%%%%%%%%%%%%%%%%%%%%%%%%%%
%%%%%%%%%%%%%%%%%%%%%%%%%%%%%%%%%%%%%%%%%%%%%%%%

      In the proton subsystem, the $ph$ excitations of interest can be divided into three 
groups, i.e., into "bound $\rightarrow$ bound", "bound $\rightarrow$ quasi-bound", and 
"bound $\rightarrow$ continuum" ones. The first group corresponds to the $ph$ excitations
from bound states located below the Fermi level into bound states located above the
Fermi one (i.e. to the excitations from the single-particle states located in the bottom
cyan shaded area into those sitting in the upper cyan shaded area [see Fig.\ \ref{fig-schematic}(a)]). 
Note that the energies of the bound single-particle states are barely affected by the change
of $N_F$ from 20 to 50 (see Sec.\ \ref{trunc-basis-sp}). In the "bound $\rightarrow$ quasi-bound" 
group, the $ph$ excitations proceed from bound single-particle states located 
below the Fermi energy [bottom cyan shaded area in Fig.\ \ref{fig-schematic}(a)]
into quasi-bound positive energy proton states located below the energy of the 
Coulomb barrier [yellow shaded area in Fig.\ \ref{fig-schematic}(a)].  Note that
the energies of quasi-bound states depend on the size of $N_F$ (see Sec.\ 
\ref{trunc-basis-sp}).  Finally, the $ph$ excitations from bound states located below the 
Fermi level into continuum states [i.e., the excitations from the single-particle states 
located in the bottom cyan shaded area into the ones sitting in the pink shaded area [see Fig.\ 
\ref{fig-schematic}(a)] build the "bound $\rightarrow$ continuum" group. According to 
Sec.\ \ref{trunc-basis-sp} the energies of continuum states are most affected by the increase
of $N_F$ (see Sec.\ \ref{trunc-basis-sp}). Note that "bound $\rightarrow$ quasi-bound" group of the $ph$ 
excitations are absent in the neutron subsystem since the $ph$ excitations in this subsystem
are built solely by the "bound $\rightarrow$ bound" and  "bound $\rightarrow$ continuum" ones 
(see bottom panels in Fig.\ \ref{fig-schematic}).  

\begin{figure}[htb]
	\includegraphics[scale=0.35]{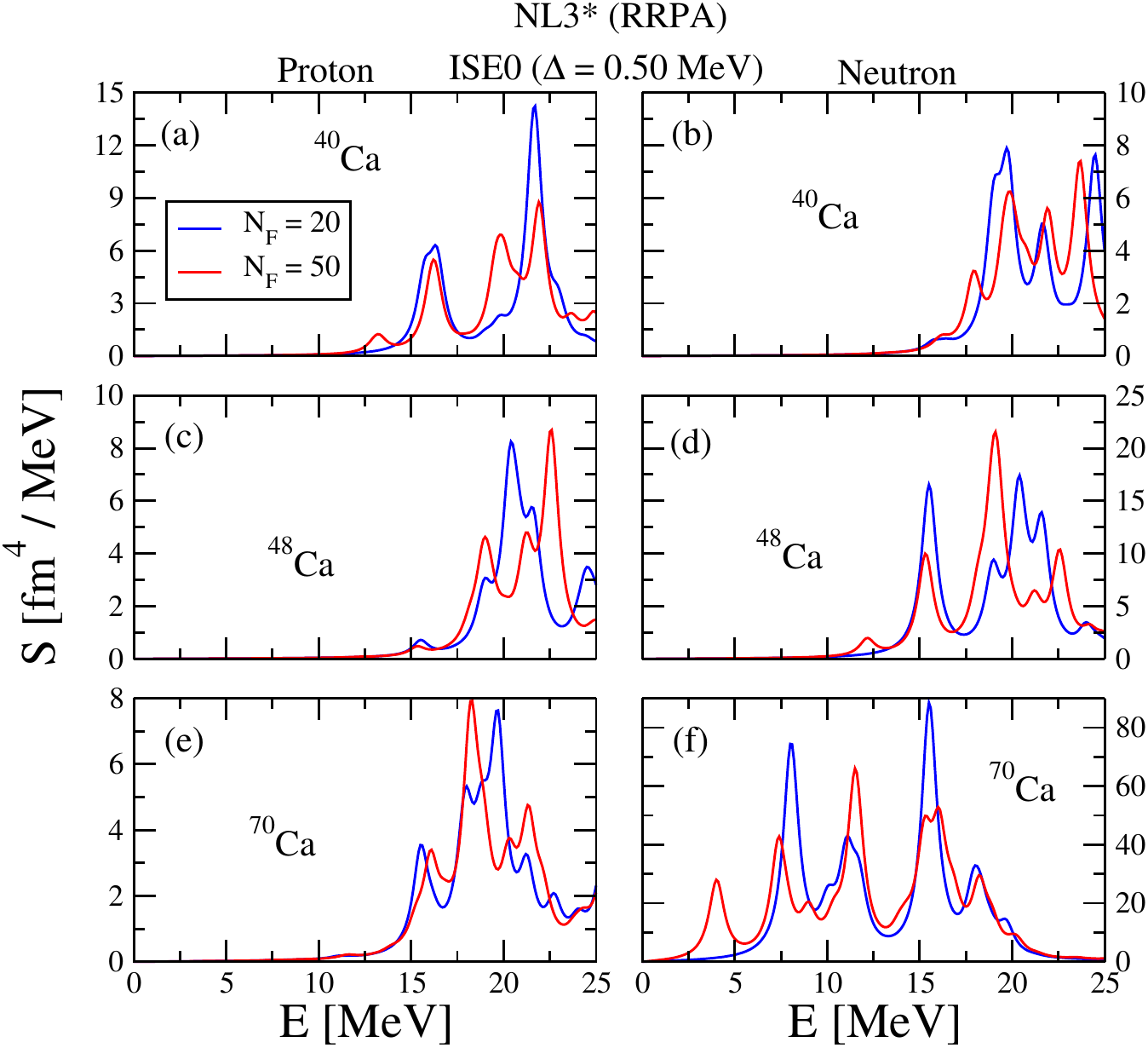}
	\caption{The dependence of proton and neutron contributions to ISGMR
	 strength function $S(E)$ on the $N_F$ value in the $^{40,48,70}$Ca nuclei. Note 
	 that the ranges on the vertical axis are different for different panels.              
\label{Prot-neutr-contr-ISGMR}
}
\end{figure}

\begin{figure}[htb]
	\includegraphics[scale=0.35]{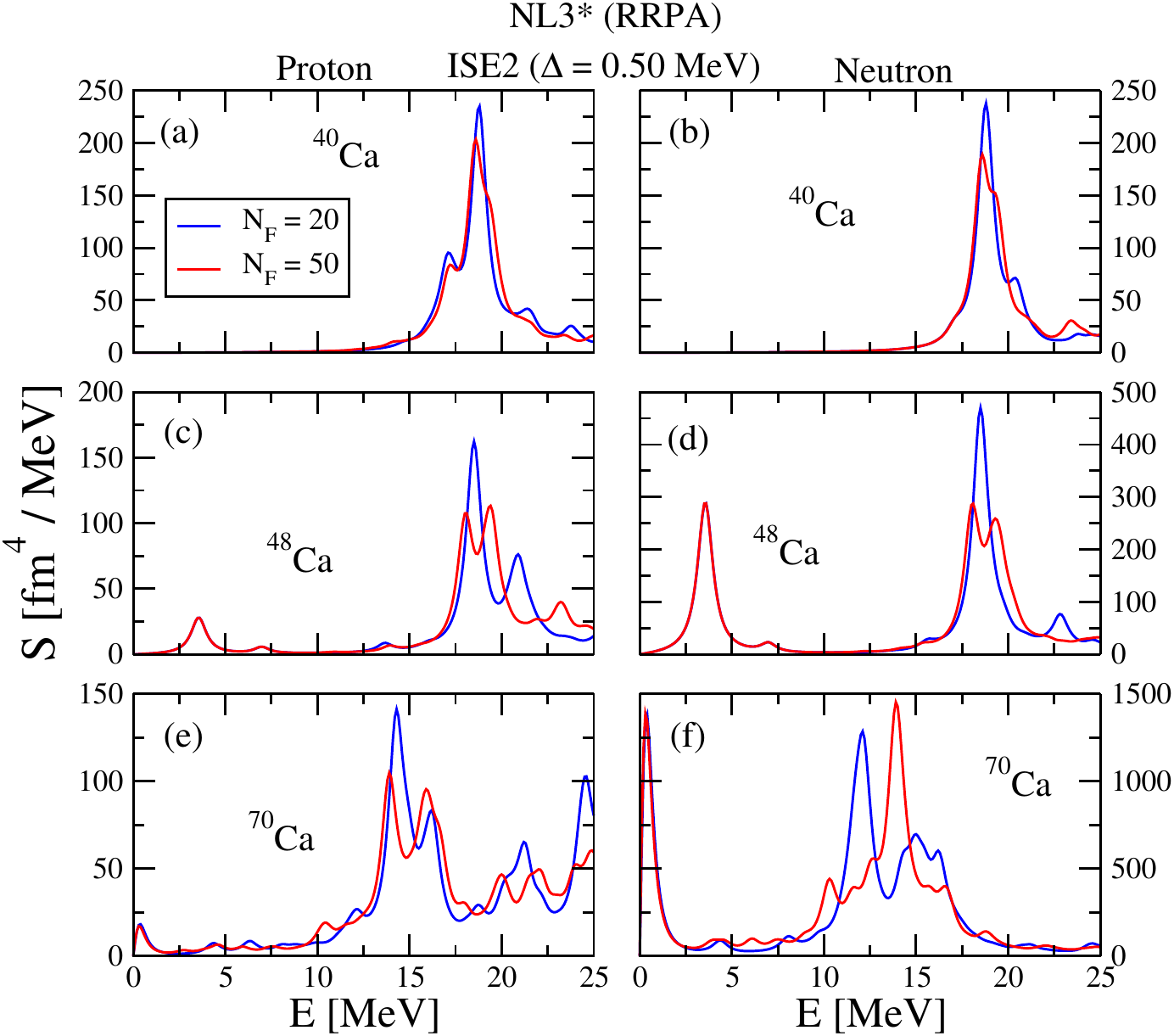}
 	\caption{The same as Fig.\ \ref{Prot-neutr-contr-ISGMR} but for the
	ISGQR.
\label{Prot-neutr-contr-ISGQR}
}
\end{figure}

     The balance of the contributions of these groups of the $ph$ excitations into the strength 
function $S(E)$ defines the impact of the increase of $N_F$ on $S(E)$. This is because
only the "bound $\rightarrow$ quasi-bound" and, especially,  "bound $\rightarrow$ continuum"
$ph$ excitations are expected to be strongly affected by the $N_F$ increase.  Moreover, 
this balance is expected to be affected by the evolution of the positions of the proton and 
neutron Fermi levels with the increase of the neutron and proton numbers. To illustrate these 
features, we analyze proton and neutron contributions to ISE0 and ISE2 nuclear response
of the $^{40,48,70}$Ca nuclei in Figs.\ \ref{Prot-neutr-contr-ISGMR}  and 
\ref{Prot-neutr-contr-ISGQR}. Note that the $^{40}$Ca nucleus is added to already discussed 
$^{48,70}$Ca nuclei since it represents an example of the $N=Z$ system. 

The individual proton and neutron contributions were obtained by retaining only the particle-hole 
pairs \{12\} of the given isospin in Eq.\ (\ref{Frho}).  Note that since $S(E)$ contains 
the squares of the respective matrix elements, the partial strength functions do not directly add 
up to the total strength function.

Let us first discuss the ISE0 response.
Figs.\ \ref{Prot-neutr-contr-ISGMR}(a) and (b) show that in the $^{40}$Ca nucleus, 
double and triple peaks appear in neutron and proton responses, respectively,  in the calculations 
with  $N_F=20$. The increase of $N_F$ to 50 leads to additional fragmentation in both responses, 
with the appearance of additional peaks and shifts of peak energies.
Note that all these changes appear in the $E\geq 15$ MeV window, which is larger than the energy 
distance between the neutron Fermi level and neutron continuum and the proton Fermi level and 
the height of Coulomb barrier (i.e. $V_{Coul}-\lambda_p$), see Table \ref{Table-sel}. This suggests that
a substantial contribution to the buildup of respective strength  functions comes from the $ph$
excitations of the "bound $\rightarrow$ quasi-bound" and "bound $\rightarrow$ continuum"  
types. Note that comparable effects are also seen in 
$^{48}$Ca [see Figs.\ \ref{Prot-neutr-contr-ISGMR}(c) and (d)].
   
%%%%%%%%%%%%%%%%%%%%%%%%%%%%%%%%%%%%%%%%%%%%%
%%%%%%%%%%%%%%%%%%%%%%%%%%%%%%%%%%%%%%%%%%%%%%%%
 
 %%%%%%%%%%%%%%%%%%%%%%%%%%%%%%%%%%%%%%%%%%%%%
%%%%%%%%%%%%%%%%%%%%%%%%%%%%%%%%%%%%%%%%%%%%%%%%
   
  The transition from $^{48}$Ca to $^{70}$Ca triggers the lowering of the proton Fermi level
from -15.6 MeV to -29.4 MeV (see Table \ref{Table-sel}).  As a result, it is reasonable to
expect the reduction of the contributions of the "bound $\rightarrow$ quasi-bound" and
"bound $\rightarrow$ continuum" $ph$ excitations into the buildup of proton $S(E)$. Indeed, 
Fig.\ \ref{Prot-neutr-contr-ISGMR}(e) shows a relatively modest difference between proton 
$S(E)$ functions calculated with $N_F=20$ and 50. This change of proton Fermi level also leads 
to a reduction of phase space for proton $ph$ excitations. As a result, the averaged proton strength 
function is approximately an order of magnitude smaller than the neutron one [compare  
Figs.\ \ref{Prot-neutr-contr-ISGMR}(e) and (d)]. This is also a reason why total
ISE0 strength is almost entirely defined by neutron response  [compare Fig.\ \ref{Strength-ISGMR}(d) 
with Fig.\ \ref{Prot-neutr-contr-ISGMR}(f)]. 
 In contrast, they have similar values in 
$^{40}$Ca  [compare  Figs.\ \ref{Prot-neutr-contr-ISGMR}(a) and (b)]. This reduction of
proton $S(E)$ function with respect to the neutron one is also seen in $^{48}$Ca [compare 
Figs.\ \ref{Prot-neutr-contr-ISGMR}(c) and (d)], but the difference is only approximately
a factor of 2.

   The transition from $^{48}$Ca to $^{70}$Ca also leads to a situation in which the
energy gap between bound and continuum neutron states reduces to almost zero
(see bottom panels of  Fig.\ \ref{fig-schematic} and Table \ref{Table-sel}). 
This has a number of important consequences. 
 First, it leads to a shift of the neutron response to substantially 
lower energies as compared with the case of $^{48}$Ca 
and to the appearance of additional peaks in the strength function [compare Figs.\ 
\ref{Prot-neutr-contr-ISGMR}(f) and (d)]. 
This is a consequence of the fact that, with approaching the neutron drip line, the role 
of "bound $\rightarrow$ bound"  $ph$ transitions decreases while that of the 
"bound $\rightarrow$ continuum"  ones drastically increase. 
  Second, the reduction of the energy gap between bound and continuum 
neutron states leads to an increase of the phase space available for  
"bound $\rightarrow$ continuum" $ph$ excitations.  This, in turn, leads to an 
increase of the average neutron strength function with increasing  the 
neutron number $N$ [compare Figs.\ \ref{Prot-neutr-contr-ISGMR}(d) and (b) 
and Figs.\ \ref{Prot-neutr-contr-ISGMR}(f) and (d)].  Third, the increase
of basis size from $N_F=20$ to $N_F=50$ modifies the magnitudes of the strength
function peaks and leads to the appearance of a soft mode at $E\approx 4$ MeV
[see Fig.\ \ref{Prot-neutr-contr-ISGMR}(f)].

     Similar to above discussed features are also seen for proton and neutron ISE2
responses (see Fig.\ \ref{Prot-neutr-contr-ISGQR}). However, the differences 
between the $N_F=20$ and $N_F=50$ results are relatively
small in the $^{40,48}$Ca 
nuclei but they become more substantial in $^{70}$Ca.  Neutron and proton strength
functions have comparable magnitudes in $^{40}$Ca. However,  with increasing neutron 
number  $N$, the proton  (neutron) strength function decreases (increases) so that the 
neutron  contribution becomes dominant in $^{70}$Ca so that it almost
entirely defines the total strength function in this nucleus 
 [compare Fig.\ \ref{Strength-ISGQR}(d) 
with Fig.\ \ref{Prot-neutr-contr-ISGQR}(f)].

    It is typically considered that the impact of the continuum on the nuclear response 
reduces with increasing mass number and the results discussed in Sec.\ \ref{RRPA-part}
and \ref{RTBA-part} for $^{132}$Sn and $^{208}$Pb generally support this picture for the 
nuclei located along the $\beta$-stability line or neutron-rich ones (such as
$^{132}$Sn, see Table \ref{Table-sel}) with proton and neutron chemical potentials 
located far below the continuum threshold. However,
the present microscopic analysis suggests that it is rather the closeness of the 
neutron Fermi level to the continuum threshold, which defines the impact of the neutron
continuum on nuclear response. This analysis suggests that such an impact is also substantial
in medium and heavy mass nuclei located not far away from the neutron drip line.

%%%%%%%%%%%%%%%%%%%%%%%%%%%%%%%%%%%%%%%%%%%%%
\begin{figure}[htb]
	\includegraphics[scale=0.23]{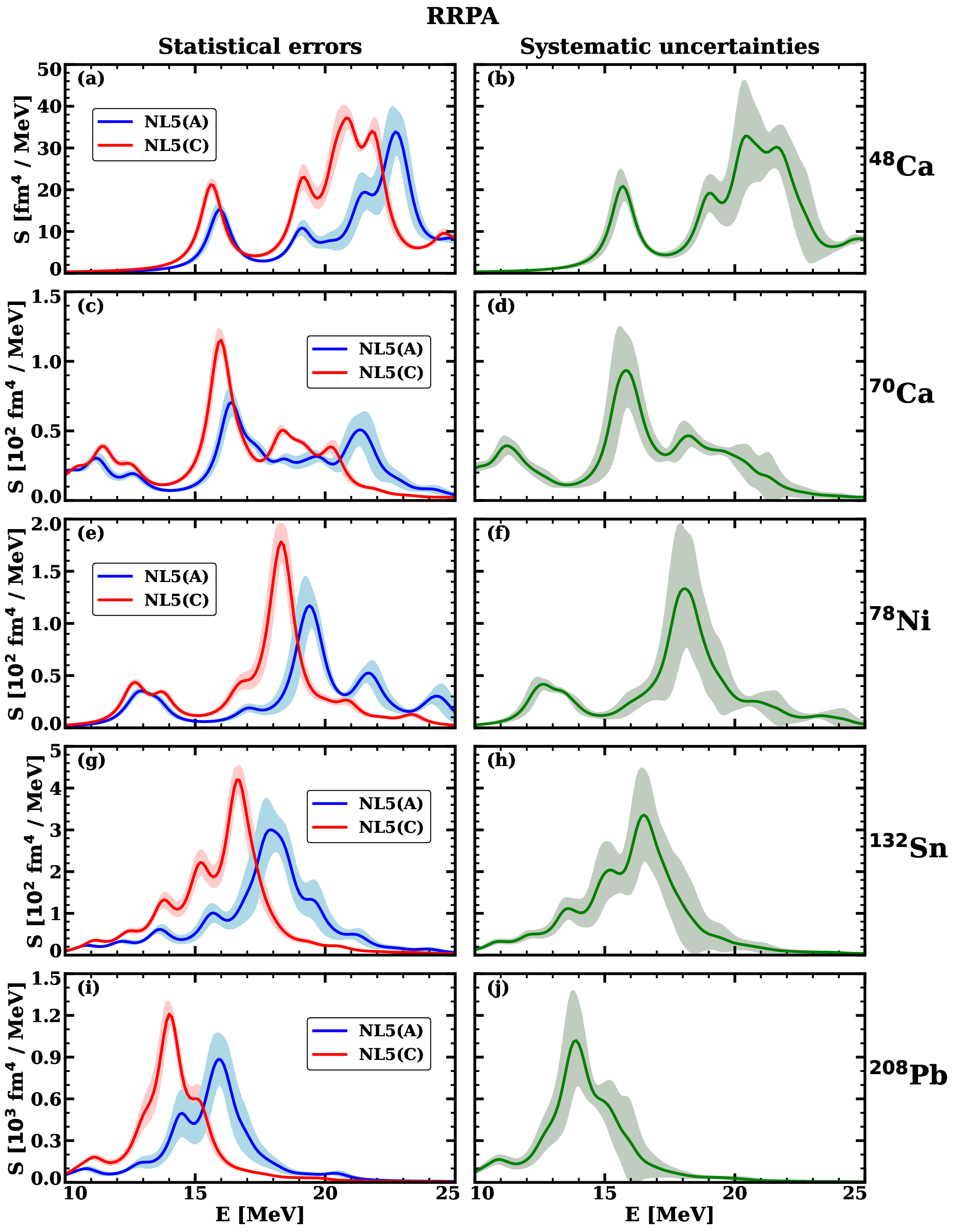} 	
 	\caption{ 
	Statistical errors (left column) versus systematic uncertainties (right column)
	of calculated ISE0 strength functions $S(E)$ in indicated  nuclei. They are shown as 
	shaded areas. The mean  $S(E)$ curves are shown by solid lines. 
	Note that the range of variables on the vertical axis depends 
	on the nucleus.
\label{stat-errors-ISE0}
}
\end{figure}
%%%%%%%%%%%%%%%%%%%%%%%%%%%%%%%%%%%%%%%%%%%%%%%%
  
%%%%%%%%%%%%%%%%%%%%%%%%%%%%%%%%%%%%%%%%%%%%%
\begin{figure}[htb]
\vspace{-0.3 cm}
	\includegraphics[scale=0.33]{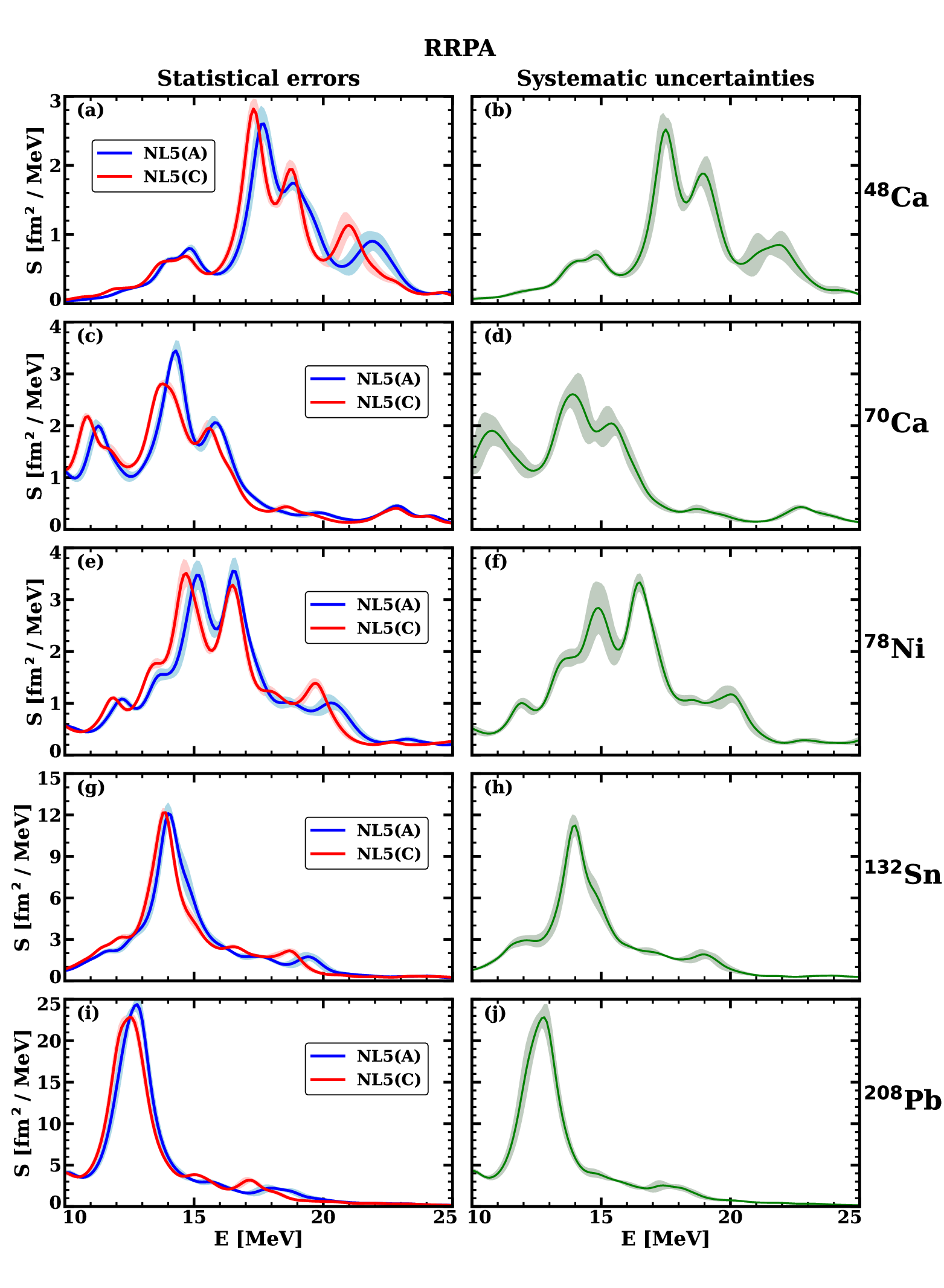} 	
 	\caption{The same as in Fig.\ \ref{stat-errors-ISE0} but for the IVE1 strength functions $S(E)$.	
\label{stat-errors-IVE1}
}
\end{figure}
%%%%%%%%%%%%%%%%%%%%%%%%%%%%%%%%%%%%%%%%%%%%%%%%

%%%%%%%%%%%%%%%%%%%%%%%%%%%%%%%%%%%%%%%%%%%%%
\begin{figure*}[htb]
	\includegraphics[scale=0.26]{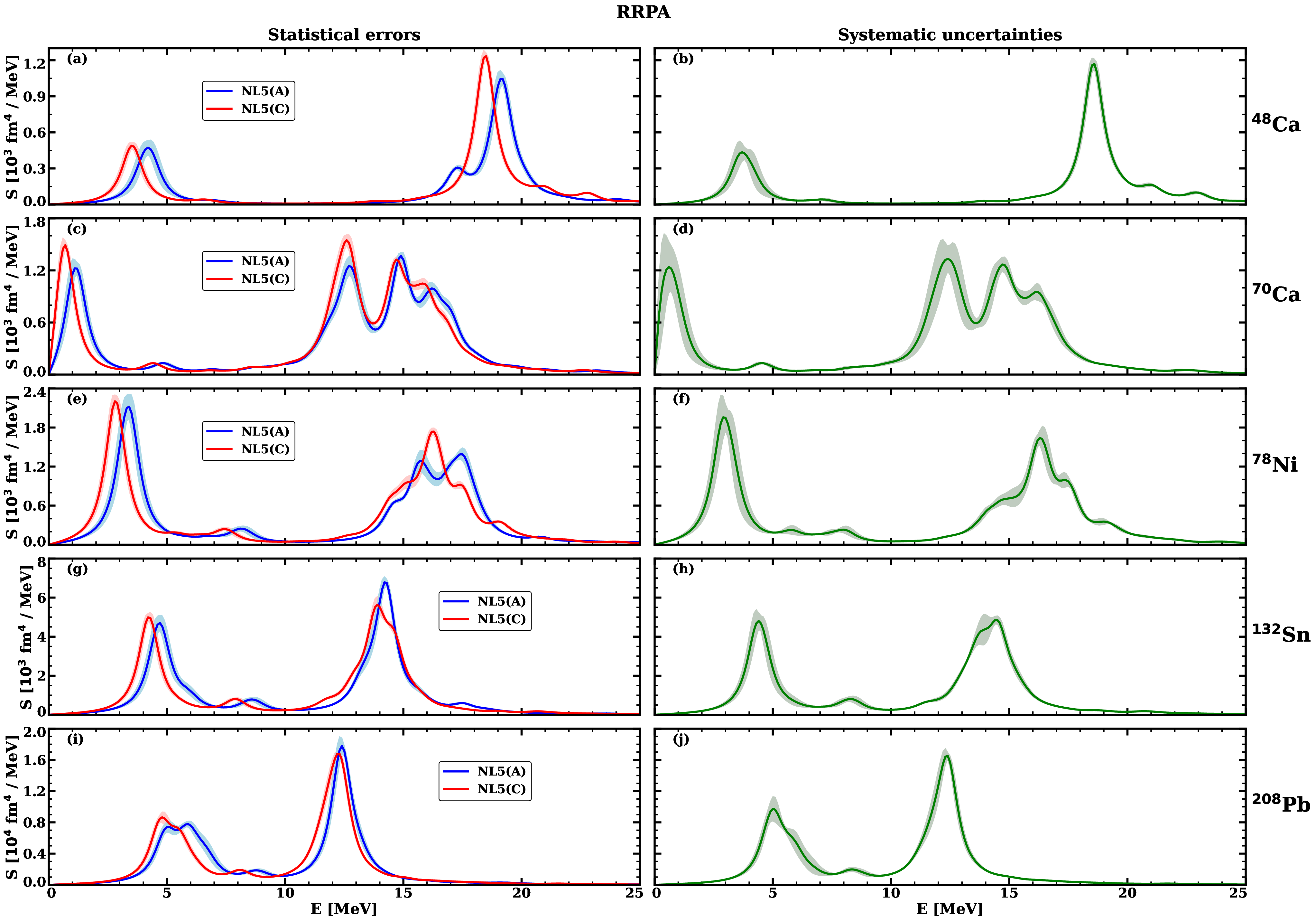} 		
 	\caption{The same as in Fig.\ \ref{stat-errors-ISE0} but for ISE2 strength functions $S(E)$.
	 For simplicity of comparison 
	  we keep the same energy unit length on horizontal
	 axis as in Figs.\ \ref{stat-errors-ISE0} and \ref{stat-errors-IVE1}: this leads to an 
	 elongation of the panels in the present figure in horizontal direction as compared with 
	 those in above mentioned figures.  
\label{stat-errors-ISE2}
}
\end{figure*}
%%%%%%%%%%%%%%%%%%%%%%%%%%%%%%%%%%%%%%%%%%%%%%%%

%%%%%%%%%%%%%%%%%%%%%%%%%%%%%%%%%%%%%%%%%%%%%
\begin{figure*}[htb]
	\includegraphics[scale=0.46]{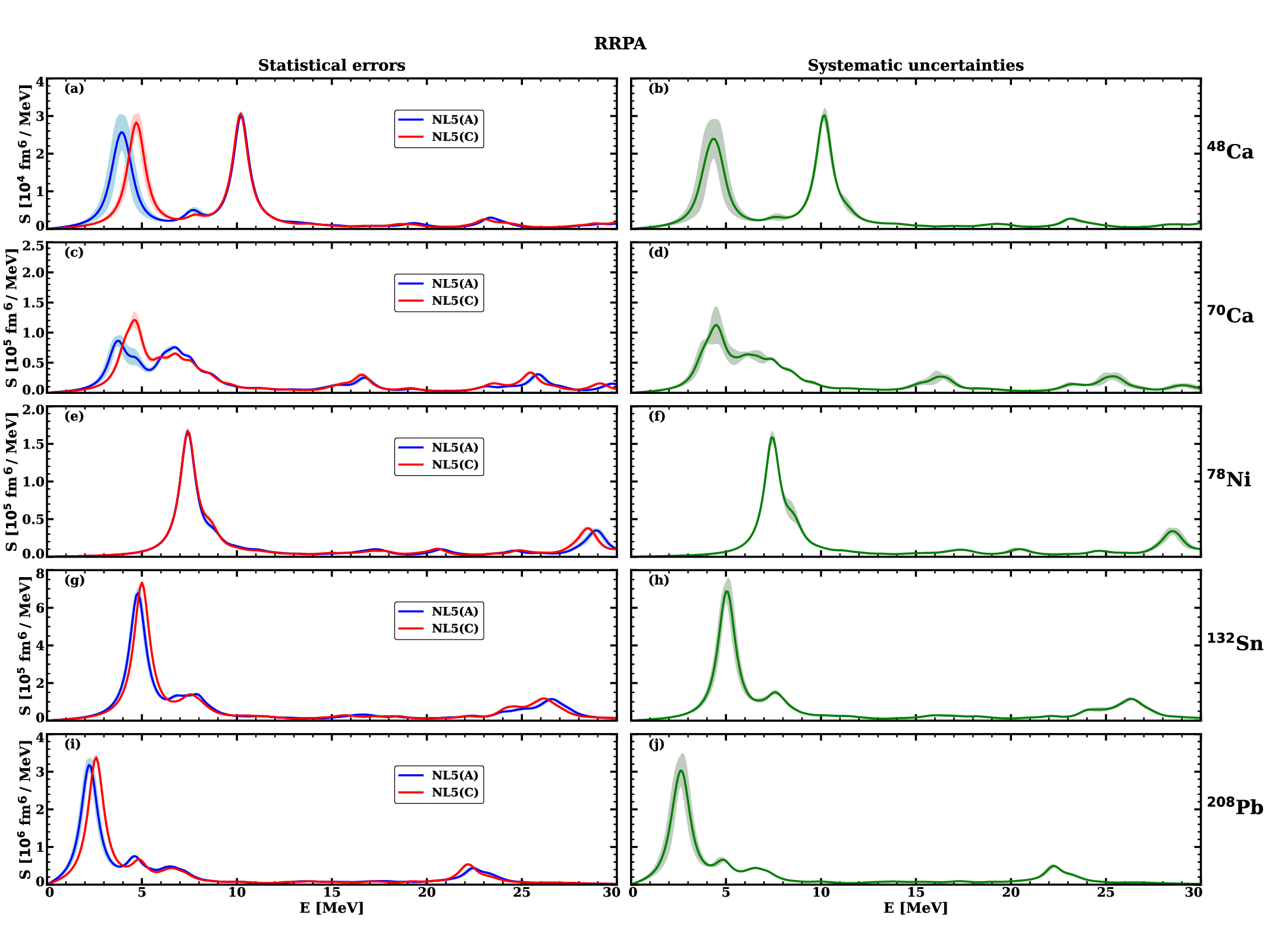} 
 	\caption{
	The same as in Fig.\ \ref{stat-errors-ISE2} but for the ISE3 strength functions $S(E)$.
\label{stat-errors-ISE3}
}
\end{figure*}
%%%%%%%%%%%%%%%%%%%%%%%%%%%%%%%%%%%%%%%%%%%%%%%%

%%%%%%%%%%%%%%%%%%%%%%%%%%%%%%%%%%%%%%%%%%%%
\section{Statistical errors and systematic uncertainties in the calculations of nuclear 
              response}
\label{statistical-errors}
%%%%%%%%%%%%%%%%%%%%%%%%%%%%%%%%%%%%%%%%%%%%

   Systematic theoretical uncertainties emerge from underlying theoretical approximations.  
In the framework of density functional theory (DFT), there are two major sources of these 
approximations, namely, the range of interaction and the form of the density dependence 
of the effective interaction \cite{BHP.03,BB.77}. Systematic uncertainties are also affected 
by the differences in the fitting protocol  details between the compared EDFs, such as the 
selection of the nuclei, the physical observables, or the corresponding weights in the 
protocols.  Because of the biases in the definition of the range of interaction and its density 
dependence as well as the  details of fitting protocols, existing EDFs do not form a 
statistical ensemble.  As a result, a systematic error  cannot be defined in a strict statistical 
sense (see Refs.\ \cite{DNR.14,AARR.14})  and instead {\it systematic 
uncertainties} in the prediction of physical observables are typically defined from the spread
of predictions given by a restricted set of functionals \cite{Eet.12,AARR.14,AAT.19}.

   In contrast, for a functional fitted in well specified fitting protocol {\it statistical errors} 
can be defined in a true statistical sense \cite{DNR.14,Stat-an,AAT.19}). They depend on the
choice of experimental data and selection of adopted errors. Note that the selection of 
adopted errors is to a degree subjective, in particular, if one deals with quantities of different
dimensions.   Finite values of  adopted errors lead to the fact that not only optimum
functional  ${\bf p}_0$ (i.e. the one corresponding to a minimum  of normalized objective function 
$\chi^2_{norm}({\bf p})$ of a given fitting  protocol) but also some functional variations
${\bf p}_i$ located in the vicinity of ${\bf p}_0$ have to be considered as acceptable. 
In a statistical sense, the variation  ${\bf p}_i$ should be considered as good as 
optimum functional ${\bf p}_0$  \cite{DNR.14,Stat-an}.

   To our knowledge, the existing calculations of nuclear response have been
performed only for optimum functionals ${\bf p}_0$  and no analysis of statistical 
errors in nuclear response 
has been carried so far.  Moreover, the same is true for systematic uncertainties 
in the  calculation of nuclear response: at present, no such analysis
has been published.
 
To fill this gap in our knowledge, we carry out a detailed analysis of statistical errors  
of strength functions $S(E)$ of different resonances in the studied nuclei. In statistical 
theory, the acceptable functionals generated during the fit are defined from the condition  
\cite{Stat-an,DNR.14}
\begin{eqnarray}
\chi^2_{norm} ({\bf p}) \leq \chi^2_{norm}({\bf p}_0) + 1 .
\label{cond}
\end{eqnarray} 
This condition specifies the 'physically reasonable' domain around ${\bf p}_0$ in which the 
parametrization ${\bf p}$ provides a reasonable fit and thus can be considered 
as acceptable. 

 The normalized objective function for model having $N_{par}$ adjustable 
parameters ${\bf p}=(p_1, p_2, ..., p_{N_{par}})$  is defined  in the fitting
protocol as
\begin{eqnarray}
\chi^2_{norm}({\bf p})=\frac{1}{s}\sum_{i=1}^{N_{type}} \sum_{j=1}^{n_i} \left( \frac{O_{i,j}({\bf p})-O_{i,j}^{exp}}
{\Delta O_{i,j}} \right)^2
\label{Ksi}
\end{eqnarray}
where 
\begin{eqnarray}
s=\frac{\chi^2({\bf p}_0)}{N_{data}-N_{par}}
\end{eqnarray}
is the global scale factor (Birge factor) \cite{Birge.32} 
defined at the minimum of the 
penalty function (optimum parametrization 
${\bf p}_0$\footnote{Because of the experimental errors and incompleteness of the 
physical modelling optimum parametrizations of the models are known 
only up to their uncertainty probability distributions \cite{DD.17}.})
which leads to the average $\chi^2 ({\bf p}_0)$ per degree of freedom 
equal to one \cite{DNR.14} and
\begin{eqnarray}
N_{data}= \sum_{i=1}^{N_{type}}n_i
\end{eqnarray}
is the total number of data points of different types. 
Here,  $N_{type}$ stands for the number of different data types. 
The calculated and experimental/empirical values of physical 
observable $j$ of the $i-$th type are represented by $O_{i,j}({\bf p})$ 
and $O^{exp}_{i,j}$, respectively. $\Delta O_{i,j}$ is adopted error 
for physical observable $O_{i,j}$.

  Based on the set of $M$ accepted functional variations 
$[{\bf p}_1, {\bf p}_2, ..., {\bf p}_M ]$  the mean values of physical 
observables 
\begin{eqnarray}
\bar{O}_{i,j}=\frac{1}{M} \sum_{k=1}^{M} O_{i,j}({\bf p}_k)
\label{mean}
\end{eqnarray}
and their standard deviations
\begin{eqnarray}
\sigma_{i,j} = \sqrt{ \frac{1}{M}  \sum_{k=1}^{M} [O_{i,j}({\bf p}_k) - \bar{O}_{i,j}]^2}
\label{st-dev}
\end{eqnarray}
are calculated for physical observables of interest. The latter serves as a measure 
of statistical error.

     The generation of the set $[{\bf p}_1, {\bf p}_2, ..., {\bf p}_M ]$ of the $M$
accepted functional variations is extremely  numerically time-consuming since 
it is performed in the $N_{par}$-dimensional  parameter hyperspace in which only 
tiny portion of randomly  generated functional variations satisfy the condition of 
Eq.\ (\ref{cond}) (see Ref.\ \cite{AAT.19}). Thus, in the assessment of statistical 
errors in nuclear response, we employ the NL5(A) and NL5(C) 
functionals for which $M=150$ and $M=500$ 
accepted functional variations were generated in Ref.\ \cite{AAT.19},
respectively (see Figs.\ 5
and 6 in this paper for the distribution of the functionals in the parameter 
hyperspace).  These functionals have a lot of similarities with the NL3* one (see Table 1 
of Ref.\  \cite{AAT.19}).

  The analysis of statistical errors in nuclear response is based here
only on the RRPA calculations, since such calculations have to be repeated
$M$ times to obtain mean values $\bar{O}_{i,j}$ and standard deviations
$\sigma_{i,j}$.  The RTBA calculations are drastically more time consuming 
than the RRPA ones. Thus, at present, the analysis of such errors cannot be 
performed in the RTBA because of numerical constraints. However, RRPA is 
the basis of RTBA, and thus it is reasonable to expect that statistical errors 
and systematic uncertainties in nuclear response are comparable 
in both approaches. Note that the RRPA
calculations presented in this Section are carried out in the $N_F=20$ basis.

  It is important to understand how large are statistical errors relative to
systematic uncertainties. The analysis of the latter is restricted to the non-linear
meson exchange functionals  and these uncertainties are obtained using
Eqs.\ (\ref{mean})  and (\ref{st-dev}) from calculated response functions defined 
for the NL3 \cite{NL3}, NL3* \cite{NL3*}, NL5(A), NL5(B), NL5(C), NL5(D), 
NL5(E) \cite{AAT.19}, NL5(Y) \cite{TA.23} and NL5(Z) \cite{NL5Z-DDMEZ-PCZ}
CEDFs\footnote{Because of these limitations of the analysis, obtained systematic
uncertainties should be considered as a lower estimate since both the addition 
of density dependent meson exchange and point coupling classes of the functionals 
and the increase of the number of the functionals in the analysis will likely to increase
these uncertainties.}. The first seven functionals are fitted to only 12 spherical
nuclei while the last two ones are defined by a global fit to the ground state
properties of all experimentally measured even-even nuclei.  Moreover, 
empirical nuclear matter properties (NMPs) are used in the fitting protocols of first 
eight functionals. In contrast, the NL5(Z) functional is fitted without information
on NMPs but in the infinite fermionic basis and with proper treatment of
total electron binding energies.  Further details on the differences of these
functionals can be found in Refs.\ \cite{NL3,NL3*,AAT.19,TA.23,NL5Z-DDMEZ-PCZ}.

  Figs.\ \ref{stat-errors-ISE0}, \ref{stat-errors-IVE1},  \ref{stat-errors-ISE2}, and
\ref{stat-errors-ISE3} compare statistical errors and systematic uncertainties 
for different responses in the  $^{48,70}$Ca,  $^{78}$Ni, $^{132}$Sn, and $^{208}$Pb
nuclei.  There are several important conclusions that follow from this comparison.

    First, statistical errors and systematic uncertainties show substantial 
dependence on the energy $E$: they are especially pronounced for energy ranges 
in which giant resonances and soft low-energy modes are located.

    Second, the largest systematic uncertainties appear for the ISE0 response
[see Figs.\ \ref{stat-errors-ISE0}(b), (d), (f), (h) and (j)].  The magnitude of 
these uncertainties does not depend significantly on the mass of the nucleus. Systematic 
uncertainties are substantially smaller for the IVE1, ISE2, and ISE3 responses
for which they have on average comparable magnitudes in the energy ranges
corresponding to giant resonances and low-energy soft modes
[compare Figs.\ 
\ref{stat-errors-IVE1},  \ref{stat-errors-ISE2} and \ref{stat-errors-ISE3}
 with Fig. \ref{stat-errors-ISE0}]. 
Statistical errors for a given functional follow the same pattern: they are 
the largest for the ISE0 response and comparable for the IVE1, ISE2 and
ISE3 ones. In addition, they are typically lower than systematic uncertainties: 
this is especially pronounced for the case of  the NL5(C) functional.  In contrast,  
statistical errors for the NL5(A) one are closer to systematic  uncertainties 
being comparable with them in some energy ranges.  
The magnitude of both statistical errors and systematic uncertainties
for a response of a given multipolarity do
not show  a pronounced dependence on mass of the nucleus.  
The substantial magnitude of statistical errors and systematic uncertainties
in the ISE0 strength function and its centroid  means that these errors and
uncertainties  have to be taken into account when 
attempting to extract incompressibility from ISGMR.

Third, one can see that typically statistical errors are by a factor of approximately
2 larger for the NL5(A) functional as compared with the NL5(C) one. This difference
is not related to the fact that the $M$ value for the NL5(A) functional is lower than the
one for  NL5(C) by a factor larger than 3. In reality, the $M=150$ number of functional
variations define statistical errors reasonably well: this is seen when such
errors for full $M=500$ set and reduced $M=150$ subsets of NL5(C) are compared.
The fitting protocols of the NL5(C) and NL5(A) functionals are almost the same:
the only differences between them are the omission of the information on neutron skin
thickness in $^{90}$Zr in the NL5(C) protocol and four-fold  reduction of adopted 
error $\Delta K_0$ in incompressibility $K_0$ [from $\Delta K_0=25$ MeV in NL5(A) 
down to $\Delta K_0 = 6.25$ MeV in NL5(C)]
(see Table 1 in Ref.\ \cite{AAT.19}). It is the latter change which leads to a substantial 
improvement of statistical errors in nuclear response on transition from the NL5(A)
functional to the NL5(C) one.

   This analysis clearly indicates that statistical errors in nuclear response are governed
to a large degree by adopted error $\Delta K_0$ used in the fitting protocol of the
functional.  However, above discussed 
reduction in adopted error may be  too optimistic since empirically recommended 
range of $K_0=190-270$ MeV \cite{RMF-nm} is substantially  larger than $\Delta K_0=6.25$ MeV. 
In the light  of the present knowledge of $K_0$, this  suggests that  a  value of adopted error $\Delta K_0$
 larger than 6.25 MeV is more realistic. In turn, this leads to a conclusion that  statistical errors in nuclear response 
obtained with NL5(A) may be more realistic than the ones calculated with NL5(C).

    It is reasonable to expect a similar distribution of statistical errors and systematic 
uncertainties in the calculation of nuclear response of different  multipolarities 
in the RTBA framework since the RRPA serves as its basis.

%%%%%%%%%%%%%%%%%%%%%%%%%%%%%%%%%%%%%%%%
\section{CONCLUSIONS}
\label{concl}
%%%%%%%%%%%%%%%%%%%%%%%%%%%%%%%%%%%%%%%%

  We performed a systematic study of the influence of the HO basis  size on the E0, E1, E2, 
and E3 response functions of doubly-magic nuclei $^{48,70}$Ca, $^{78}$Ni, $^{132}$Sn, 
and $^{208}$Pb in wide energy intervals.  The calculations confined to the $N_F$ = 20 HO shells were confronted with 
those in a considerably extended basis of  $N_F$ = 50 HO shells executed for the first 
time in (i) RRPA and (ii) RRPA extended by the minimal particle vibration coupling dubbed 
as RTBA. 

 The increase of the HO basis leads to a lowering of the energies 
of the  single-particle states located above zero energy threshold and to an increase in the 
densities of quasi-bound and continuum single-particle states in the energy domain relevant 
to the formation of nuclear soft modes and giant resonances below 25 MeV.
This reflects the different sensitivity of the bound and positive energy single-particle states to 
the properties of the HO basis.  The convergence of bound states in energy is defined 
by nucleonic potential, i.e. it is directly related to the convergence of total binding energies
of nuclei.  In contrast, the hard-wall boundary conditions in the coordinate space effectively 
imposed by the HO  basis in nuclear many-body calculations define the convergence in
energy of positive energy states.  As a result, it is much more difficult to reach a full 
convergence in energy for quasi-bound and continuum states than for bound ones.
There are significant similarities in the evolution of the energies of positive energy
single-particle states with an increase of both the $N_F$ value in the HO basis calculations
and the radius $R_{box}$ of the spherical box in coordinate space calculations. This suggests
that the difference between these two types of calculations reduces with an increase in $N_F$.
 
These observations indicate that the errors due to the truncation of the basis considerably reduce 
with an increase in the HO basis from $N_F=20$ to $N_F=50$.
Although in this work we did not reach full convergence of positive energy single-particle states with respect 
to the basis size, this may be technically possible, and this capability will be 
investigated in the future.
The RRPA and RTBA  results reveal an enhanced sensitivity to the HO basis completeness for the 
monopole excitations and light nuclei.  However, sizeable but smaller effects are  
also seen for excitations of higher multipolarities and medium and heavy mass nuclei. 
This suggests that the truncation of fermionic HO basis introduces errors, which are
considerable in some cases, that are not of theoretical but of purely technical character.  

The statistical errors and systematic uncertainties in nuclear response have also been 
investigated.  They are especially pronounced for the E0 response, but become smaller
for higher multipolarity ones. Note that they do not show a pronounced dependence on
the mass of the nuclei. The statistical errors are typically smaller than 
systematic uncertainties, but their magnitudes sensitively depend on the adopted
error $\Delta K_0$ in incompressibility $K_0$ employed in the fitting protocol
of the functional. Although this part of the study has been carried out only in the RRPA 
framework, a similar  picture is expected for statistical errors and systematic uncertainties 
in the RTBA calculations.

The obtained results set a new perspective for theoretical studies of the nuclear 
response.  Critical characteristics, such as the nuclear compressibility, dipole polarizability, quadrupole 
instability, and octupole softness, are sensitive to the HO basis completeness and are 
affected by the statistical errors and systematic uncertainties in the calculations of nuclear
response. These factors have to be taken into account in their calculations and in benchmarking 
those properties for extrapolations to the nuclear matter equation of state.
   The continuum effects on simple and complex configurations are pronounced in fine spectral details, 
 which are especially important at low energies below and around the particle emission threshold. Although 
 further studies in the direction of more extended HO bases are desirable for the giant resonances, the 
 account for the continuum within the $N_F$ = 50 basis can already be acceptable for the soft modes or their considerable portions located below the particle emission threshold.  A complementary development capable of treating free asymptotic of the unbound nucleonic states would further shed light on the accuracy of numerical modeling of the nuclear response in the HO bases.

\section{ACKNOWLEDGMENTS}

 This material is based upon work supported by the U.S. Department of Energy,  Office of Science, 
Office of  Nuclear Physics under Award No. DE-SC0013037  and the US-NSF Grants PHY-2209376 
and PHY-2515056.

%===============================================================================
% \section*{Acknowledgement}
%
% The work of E.L. was supported by the US-NSF Grants PHY-2209376 and PHY-2515056.
%
%===============================================================================

%\bibliography{references-44-next-gen-CEDFs.bib}
\bibliography{references-48-optimized-basis.bib}
\end{document}